\documentclass[12pt]{article}
\usepackage{graphicx}
\def\ltap{\raisebox{-.4ex}{\rlap{$\sim$}} \raisebox{.4ex}{$<$}}

\def\sinthl {${\rm sin}^2 \theta_W^{\ell\ {\rm eff}}$}
\def\sinthlsp {${\rm sin}^2 \theta_W^{\ell\ {\rm eff}}\ $}
\def\qx {$Q_X$}
\def\qxsp {$Q_X\ $}
\def\zp {$Z^{\prime}$}
\def\zpsp {$Z^{\prime}\ $}
\def\zzerop {$Z^{0\, \prime }$}
\def\zzeropsp {$Z^{0\, \prime}\ $}
\def\sp {$S^{\prime}$}
\def\spsp {$S^{\prime}\ $}
\def\tp {$T^{\prime}$}
\def\tpsp {$T^{\prime}\ $}
\def\tx {$T_X$}
\def\txsp {$T_X\ $}

\def\zzpsp {$Z-Z^{\prime}\ $}
\def\hpsp {$H^{\prime}\ $}
\def\hp {$H^{\prime}$}
\def\gzpsp {$G_{Z^{\prime}}\ $}
\def\gzp {$G_{Z^{\prime}}$}
\def\ypsp {$Y^{\prime}\ $}

\def\alr {$A_{LR}$}
\def\alrsp {$A_{LR}\ $}
\def\afbb {$A_{FB}^b$}
\def\afbbsp {$A_{FB}^b\ $}
\def\afbc {$A_{FB}^c$}
\def\afbcsp {$A_{FB}^c\ $}
\def\qfb {$Q_{FB}$}
\def\qfbsp {$Q_{FB}\ $}

\def\afbl {$A_{FB}^{\ell}$}
\def\atau {$A_{\ell}(P_{\tau})$}
\def\atausp {$A_{\ell}(P_{\tau})\ $}
\def\mh {$m_H$}
\def\mhsp {$m_H$\ }
\def\chisqsp {$\chi^2$\ }
\def\chisq {$\chi^2$}

\def\mwsp {$m_W\ $}

\def\hatgzpsp {$\hat G_{Z^{\prime}}\ $}

\def\journal{\topmargin 0.0in   \oddsidemargin 0in
        \headheight 0pt \headsep 0pt
        \textwidth 6.5in 
\textheight 9in 
        \marginparwidth 1.5in
        \parindent 2em
        \parskip .5ex plus .1ex         \jot = 1.5ex}
\journal

\begin{document}
\begin{titlepage}

\noindent May 24, 2008   \\  
\noindent Revised June 24, 2008

\begin{center}

\vskip .5in

{\large A $Z^{\prime}$ Boson and the Higgs Boson Mass}

\vskip .5in

Michael S. Chanowitz

\vskip .2in

{\em Theoretical Physics Group\\
     Lawrence Berkeley National Laboratory\\
     University of California\\
     Berkeley, California 94720}
\end{center}

\vskip .25in

\begin{abstract}

The Standard Model fit prefers values of the Higgs Boson mass $m_H$
that are below the 114 GeV direct lower limit from LEP II. The
discrepancy is acute if the $3.2\sigma$ disagreement for ${\rm
sin}^2 \theta_W^{\ell\ {\rm eff}}$ from the two most precise
measurements is attributed to underestimated systematic error. In that
case the data suggests new physics to raise the predicted value of
\mh. One of the simplest possibilities is a \zpsp boson, which would
generically increase the prediction for \mhsp as a result of $Z$-\zpsp
mixing. We explore the effect of $Z$-\zpsp mixing on the \mhsp
prediction, using both the full data set and the reduced data set that
omits the hadronic asymmetry measurements of ${\rm sin}^2
\theta_W^{\ell\ {\rm eff}}$, which are more likely than the leptonic
asymmetry measurements to have underestimated systematic uncertainty.

\end{abstract}

\end{titlepage}

\newpage

\renewcommand{\thepage}{\arabic{page}}
\setcounter{page}{1}

\noindent {\bf 1. Introduction} 

The Standard Model fit of the precision electroweak data, reviewed
below, has a less than robust $\chi^2$ confidence level,
$CL(\chi^2,N)=CL(17.2,12)=0.14$, as a result of the enduring
$3.2\sigma$ discrepancy between the two most precise measurements of
the effective leptonic weak mixing angle, ${\rm sin}^2 \theta_W^{\ell\ {\rm
eff}}$, from the polarization asymmetry \alrsp and the front-back $b$
quark asymmetry \afbb. Since the SM fit is relied on to provide
guidance on the mass of the Higgs boson, it is relevant to consider
the consistency of the sector of measurements that predict the value
of $m_H$. In this sector the problem is more severe, with
$CL(\chi^2,N)=CL(14.1,7)=0.05$.  The discrepancy between \alrsp and
\afbbsp is reflected in a $3.2\sigma$ discrepancy between the three
leptonic asymmetry measurements, \alr, \afbl, \atau, and the three
hadronic asymmetry measurements, \afbb, \afbc, and \qfb, and in the
poor \chisqsp for the combination of all six asymmetries,
$CL(11.8,5)=0.037$.\cite{ewwg} These discrepancies could be
statistical fluctuations, evidence of new physics, or the result of
underestimated systematic uncertainty. If they are due to new physics,
we cannot extract the Higgs boson mass \mhsp from the precision data
without first specifying the nature of the new physics.

It might appear that the viability of the SM could be enhanced if
the discrepancies are attributed to underestimated systematic
uncertainty, in particular, in the hadronic asymmetry measurements,
which share challenging, common experimental and theoretical systematic
uncertainties. Indeed, if that is assumed and the three hadronic
asymmetry measurements are omitted from the fit, the confidence level
increases from 0.14 to 0.78, but a new problem emerges: the remaining
measurements, dominated by \alrsp, $m_W$, and $m_t$, predict $m_H=50$
GeV, with only a small probability, $CL(m_H > 114)=0.03$, for \mhsp in
the region $m_H > 114$ GeV allowed by the LEP II direct search
limit.\cite{msc123} Therefore this scenario also suggests new physics,
in this case new physics to raise the predicted value of \mh, and once
again \mhsp cannot be extracted from the data without specifying a
model for the new physics. 

With this motivation several models of new physics have been
considered to raise the the predicted value of \mhsp in the fit with
hadronic asymmetries excluded, including light sneutrinos and
gauginos,\cite{guidoetal} a fourth family of quarks and
leptons,\cite{levetal} and mixing with heavy vector-like
leptons\cite{vleptons}.  In this paper we consider mixing of the SM
$Z$ boson with a heavy \zpsp boson associated with a new Abelian
symmetry with generator \qx, a simple extension of the SM that can
raise the predicted value of \mh. The mechanism is easy to understand:
a heavy Higgs boson makes a negative contribution to the $\rho$
parameter, $\rho=m_W^2/m_Z^2{\rm cos}^2\theta_W$, while mixing of $Z$
with a heavier \zpsp shifts $m_Z$ downward, causing $\rho$ to increase
so that the two effects tend to cancel. This possibility has been
explored by Ferroglia, Lorca, and van der Bij\cite{flv} for the
reduced data set with \afbbsp excluded, for \zpsp bosons coupled to
weak hypercharge $Y$ and to $B-L$, the difference of baryon and lepton
number. Our results agree qualitatively with theirs but differ in
detail, both in the formulation of the \zpsp model and in the
implementation of the experimental constraints. In our approach but
not in theirs the $Z$-\zpsp mass matrix is generated by Higgs bosons
in the conventional way.\footnote{The authors of \cite{flv} exploit
  the fact that the Higgs mechanism is not necessary to ensure
  renormalizability in the case of Abelian gauge bosons.} This
theoretical difference has experimental consequences which are
discussed below. They fit a truncated data set that captures the
principal features but differs in detail from our fits, which are
based on the complete EWWG\cite{ewwg} data set, use
ZFITTER\cite{zfitter} to compute the radiative corrections, and
include the largest experimental correlations as given by the EWWG.
In addition, we impose the constraints on \zpsp production extracted
by Carena {\it et al.}\cite{cddt} from the LEP II bounds for BSM
contact interactions,\cite{lep2} which we find are stronger than the
precision EW constraints in parts of the \qxsp parameter space. We
also impose the more recent constraints on \zpsp models obtained by
the CDF collaboration,\cite{cdf} which are stronger than the LEP II
bounds for some of the \qxsp parameter space if the \zpsp coupling
constant is sufficiently small,\footnote{I thank Bogdan Dobrescu for
  bringing the CDF bounds to my attention.} in particular, smaller
than electroweak strength. Fits both with and without the hadronic
asymmetries are presented.

Following \cite{ceg,adh} we consider a class of models in which (1)
the \zpsp receives its mass from a heavy SM singlet Higgs boson \hp, (2)
the new gauge group $U(1)_X$ is required to be anomaly free with
matter fields restricted to three SM generations augmented only by
three right-handed neutrinos, and (3) the \qxsp charges of the SM
fermions are independent of generation. It then follows\cite{ceg,adh}
that \qxsp must act on quarks and leptons like a linear combination of
SM hypercharge $Y$ and $B-L$, say\footnote{ A peculiar third solution
  obtained in \cite{ceg} is incorrect, because those authors
  apparently failed to consider the $SU(3)_C^2 \times U(1)_X$
  anomaly.}
$$
Q_X= {\rm cos}\theta_X {Y \over 2} + {\rm sin}\theta_X {B-L \over 2}. 
                                          \eqno{(1)}
$$ 

The SM fermions and the $W^{\pm},Z_0$ bosons obtain their masses from
the usual SM Higgs boson, which to preserve \qxsp gauge invariance
must also be assigned \qxsp charge ${\rm cos}\theta_X {Y \over 2} +
{\rm sin}\theta_X {B-L \over 2} $ with its usual SM $Y$ and
(vanishing) $B-L$ charges. In this approach, in order for there to be
$Z$-\zpsp mass mixing, the SM Higgs boson must have nonvanishing \qxsp
charge, $Q_X H \neq 0$, requiring $|\theta_X| \neq \pi/2$. This
contrasts with the model of \cite{flv} in which $Z$-\zpsp mixing can
occur even if $Q_X$ acts on SM quanta purely like $B-L$. Following
\cite{adh}, we use the freedom to define the SM $B$-hypercharge gauge
boson and the new singlet \zpsp so that kinetic mixing vanishes at the
electroweak scale and $Z$-\zpsp mixing is completely described by the
mass matrix for the relevant energies near the TeV scale.  We assume
the SM singlet Higgs boson \hpsp has a very large vacuum expectation
value, $v^{\prime}\gg v$, and that the new vector boson is much
heavier than the $Z$, $m_{Z^{\prime}} \gg m_Z$.

\qxsp then coincides with the SM generators ${\rm cos}\theta_X {Y
  \over 2} + {\rm sin}\theta_X {B-L \over 2}$ in its action on SM
matter quanta but not in its action on BSM quanta such as \hp. In this
framework with $\theta_X=0$ there can be a class of ``$Y$-sequential''
\zpsp bosons with charges \ypsp which are identical to the SM
hypercharge $Y$ in their action on SM quanta but differ in their
action on BSM quanta.\cite{ceg,adh} These models were described as
``unaesthetic'' in \cite{ceg}, although with a caveat that did not
survive the journal's editorial process but is reproduced here: ``We
are humbly aware that aesthetic judgements are subjective and
time-dependent. The cockroaches of Troy probably did not understand
why the Greeks were making so much fuss.''\footnote{See the footnote
  on page 10 of the scanned preprint of \cite{ceg} posted at
  http://ccdb4fs.kek.jp/cgi-bin/img\_index?197706199.}  This class of
models has an appreciable effect on the allowed range of \mhsp in the
EW fits, and especially for the fit of the reduced data set.

The issues raised by the precision EW data will continue to be
important in the era of the LHC. Just as it has played an important
role in the development of the SM, the precision EW data can also help
us to understand the discoveries that will be made at the LHC. But our
ability to use the precision EW data for this purpose will be severely
limited if we cannot resolve the ambiguity created by the \afbbsp
anomaly.

In section 2 we review the SM fit and predictions for the Higgs boson
mass. In section 3 we describe the class of \zpsp models to be
considered. In section 4 we summarize the relevant LEP II constraints
on \zpsp bosons, taken from \cite{cddt} and \cite{lep2}, and the more recent
constraints from CDF.\cite{cdf} In section 5 we present constraints
from the fits to the precision electroweak data together with the LEP
II constraints. Concluding remarks are given in section 6. In an
appendix we show that for a $Y$-sequential \zpsp the effect of \zzpsp
mixing on the SM fit can be fully represented by ``pseudo-oblique''
corrections, which account for both vacuum polarization and vertex
corrections.

\noindent {\bf 2. The Standard Model Fit} 

In this section we review the SM fit to the precision electroweak
data. We use the data set and methodology of the EWWG\cite{ewwg} with
one exception: we do not include the $W$ boson width in our fits,
since with a 2.5\% error it is not a precision measurement in the
sense of the others, which are typically measured to O(0.1\%) or
better, and in any case it has no impact on the prediction for the
Higgs boson mass. We include the largest experimental correlations as
given by the EWWG and use ZFITTER\cite{zfitter} to compute the
radiative corrections, including the two loop contributions to
\sinthlsp amd $m_W$.\cite{twoloop} Like the EWWG we perform a \chisqsp
fit to the data, scanning over $m_t, \Delta \alpha^{(5)}_{\rm
  had}(m_Z), \alpha_S(m_Z)$, and \mh,\footnote{We have verified that
  the fit is not affected by scanning on $m_Z$ because it is much more
  precisely measured than the other observables.} leaving the latter
two parameters unconstrained. The fits use the most recent Fermilab
measurement of the top quark mass,\cite{mt} $m_t= 172.6 \pm 1.4$ GeV.

In table 1 our fit is compared with the most recent EWWG
fit\cite{last-ewwg}, where it is clear the two are virtually
indistinguishable: with $\Gamma_W$ omitted both yield $\chi^2/N=
17.2/12$ (with correlations contributing $-1.4$). The difference for
$m_t$ is an artifact of our fitting grid, $\Delta m_t = .21$ GeV,
which has been overtaken by the increasing experimental precision. The
consistency for all other quantities, one part per {\it mil} or
better, shows that the coarseness of the $m_t$ grid has not affected
the quality of the fit.  For the Higgs mass our central value is 85
GeV, indistinguishable from 87 GeV obtained by the EWWG.

\begin{table}
\begin{center}
\vskip 12pt
\begin{tabular}{c|cccc}
\hline
\hline
 &Experiment& EWWG SM Fit& Our SM Fit &Pull \\ 
\hline
\hline
$A_{LR}$ & 0.1513 (21)  & 0.1480  & 0.1480  &1.6  \\
$A_{FB}^l$ & 0.01714 (95) &0.01643  &0.01642  & 0.76 \\
$A_{e,\tau}$ & 0.1465 (32) & 0.1480 & 0.1480 & -0.45 \\
$A_{FB}^b$ & 0.0992 (16) & 0.1038 & 0.1037 &-2.8  \\
$A_{FB}^c$ & 0.0707 (35) & 0.0742 & 0.0741 & -1.0 \\
$x_W^l[Q_{FB}]$ & 0.2324 (12) & 0.2314 & 0.2314 & 0.83  \\
$m_W$ & 80.398 (25) & 80.377 & 80.374 & 0.95 \\
$\Gamma_Z$ & 2495.2 (23) & 2495.9 & 2495.9 &-0.3  \\
$R_l$ & 20.767 (25) &20.743  &20.744  &1.0  \\
$\sigma_h$ & 41.540 (37) & 41.478 & 41.477 &1.7  \\
$R_b$ & 0.21629 (66) & 0.21581  & 0.21586 &0.65  \\
$R_c$ & 0.1721 (30) & 0.1722 & 0.1722 &-0.04  \\
$A_b$ & 0.923 (20) & 0.935& 0.935  &-0.6  \\
$A_c$ & 0.670 (27) &  0.668 &  0.668 & 0.07 \\
$m_t$ & 172.6 (1.4) &172.8  &172.3  &0.24  \\
$\Delta \alpha_5(m_Z^2)$ & 0.02758 (35) &0.02767&0.02768& 0.29 \\
$\alpha_S(m_Z)$ &  &0.1185  &0.1186  &  \\
$m_H$ & & 87 & 85 & \\
\hline
\hline
\end{tabular}
\end{center}
\caption{SM fit compared with the EWWG fit.\cite{last-ewwg}}
\end{table}

We can see from table 1 that the less than robust \chisqsp confidence
level of the SM fit, $CL(17.2,12)=0.14$, is a consequence of the 3.2
$\sigma$ discrepancy between the leptonic and hadronic asymmetry
measurements. \afbbsp is the measurement with the largest pull,
$2.82\sigma$, corresponding to a a nominal Gaussian confidence level
of 0.0048. The significance of such an outlyer can be estimated by the
probability that one of twelve independent measurements will fluctuate
to $\geq 2.82\sigma$, which is $1-(1-0.0048)^{12}=0.06$, enough by
itself to account for the less than robust confidence level of the
global fit. If we consider only the observables that are sensitive to
$m_H$, omitting $R_b, R_c, A_b, A_c, \sigma_H$ which are not, the
\chisqsp confidence level falls by a factor 3 to $CL(14.1,7)= 0.05$,
and \afbbsp is again the leading outlyer with a pull of $2.80\sigma$
and nominal likelihood 0.0051. The probability for such an outlyer is
then $1-(1-0.0051)^{7}=0.035$, which matches nicely with the 0.05
confidence level of the corresponding \chisqsp fit. For both of these
fits the central value for the Higgs boson mass is $m_H=85$ GeV, but
since $CL(m_H>114)=0.26$ there is no significant conflict with the LEP
II 114 GeV lower limit.

It is instructive to consider why \afbbsp is the outlyer in these fits
rather than \alr. The explanation is that there is an ``alliance''
between the leptonic asymmetry measurements and the $W$ boson mass
against the hadronic asymmetry measurements. The former favor very low
values for the Higgs boson mass, $m_H\, \ltap\, 50$ GeV, as opposed to
the hadronic asymmetry measurements, which predict $m_H$ an order of
magnitude heavier, $m_H \simeq 500$ GeV. In table 2 the central value
from a fit to \afbbsp alone is $m_H=480$ GeV, while \alrsp and \mwsp
predict values an order of magnitude smaller, $m_H=34$ and $52$ GeV
respectively. We see in table 2 and figure 1 that these differences
are significant: the 90\% symmetric confidence intervals for \mhsp
from the leptonic asymmetries and from the nonasymmetry measurements
do not overlap the 90\% interval from the hadronic asymmetries. (Note
that the 95\% confidence level upper limit is just the upper limit 
of the symmetric 90\% confidence interval.)

\begin{table}
\begin{center}
\vskip 12pt
\begin{tabular}{c|ccc}
\hline
\hline
 & $m_H$ (GeV) & 90\% $CL$ & $CL(m_H\, >\, 114)$ \\ 
\hline
\hline
$A_{LR}$ & 34 & $ 10-<m_H<108$ & 0.07 \\
$m_W$ & 52 & $15 <m_H< 135$ & 0.10\\
$A_{FB}^b$ & 480 & $170 <m_H< 1000+$ & 0.99 \\
\hline
$A_{LR}\oplus A_{FB}^l \oplus A_{\tau}$&50&$ 19<m_H<126$ & 0.07 \\
$m_W\oplus \Gamma_Z \oplus R_l$ & 52 &$ 10 <m_H< 140$ & 0.10 \\
$A_{FB}^b \oplus A_{FB}^c \oplus Q_{FB}$ & 480&$180 <m_H< 1000+$ & 0.99 \\
\hline
\hline
\end{tabular}
\end{center}
\caption{Predictions for $m_H$ from various restricted
sets of $m_H$-sensitive observables. The value of $m_H$ at the
$\chi^2$ minimum is shown along with the symmetric 90\% confidence
interval and the likelihood for $m_H\, >\, 114$ GeV. $10-$ and $1000+$ 
denote intervals extending below 10 or above 1000 GeV.}
\end{table}

\begin{figure}[t]
\centerline{\includegraphics[width=4.2in,angle=90]{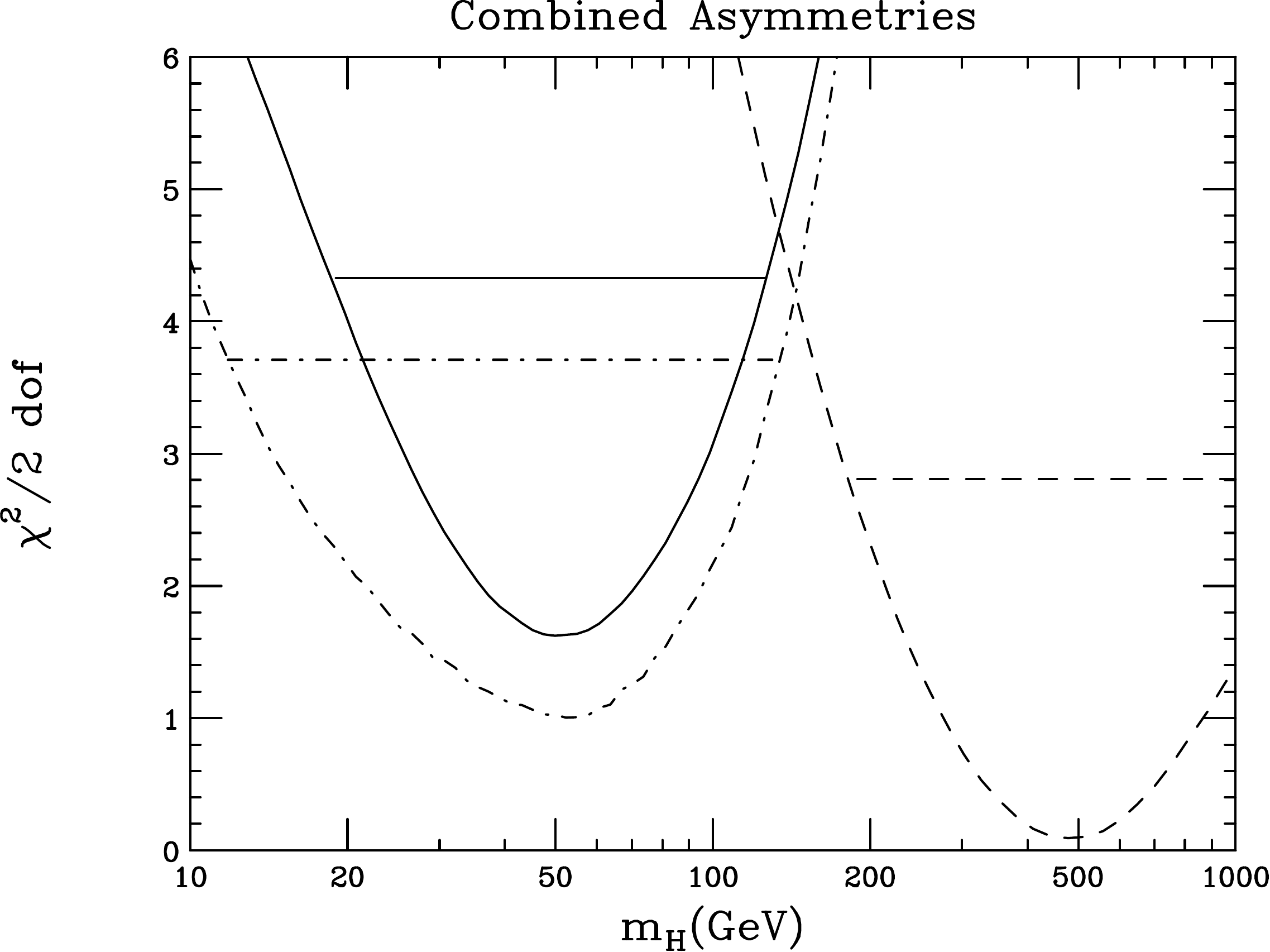}}
\caption{$\chi^2$ distributions as a function of $m_H$ from the
  combination of the three leptonic asymmetries \alr, \afbl, \atausp
  (solid line); the three hadronic asymmetries \afbb, \afbc, and
  \qfbsp (dashed line); and the three \mh-sensitive, nonasymmetry
  measurements, $m_W, \Gamma_Z$, and $R_l$ (dot-dashed line).  The
  horizontal lines indicate the respective 90\% symmetric confidence
  intervals.}
\label{fig1}
\end{figure}

The interpretation of the precision data then depends critically on
how we interpret the discrepancy between the hadronic and leptonic
asymmetry data. If it is a statistical fluctuation then the prediction
for \mhsp from the \chisqsp fit to the full data set is applicable,
with central value 85 GeV and 95\% upper limit 158 GeV. Since the
\afbbsp and \alrsp measurements both represent many years of careful
work, it is also certainly possible that the discrepancy is a genuine
reflection of new physics, for instance, in the $Z\overline bb$
vertex. Because $R_b$ agrees well with the SM prediction, it is
straightforward to show that this hypothesis requires a very large
($\sim 20\%$) new physics contribution to the right-handed $Z\overline
bb$ coupling. Popular models of new physics cannot readily explain the
data, but there is not a no-go theorem and some possibilities have
been explored.\cite{afbb-np} If it is a genuine manifestation of new
physics, then the new physics must first be known in order to use the
precision data to predict \mh.

The third possible explanation of the discrepancy is underestimated
systematic error. The three leptonic measurements are comparatively
straightforward. They are free of complications from QCD and
hadronization, involve three quite different techniques with no common
systematic uncertainties, and have a sensible \chisq, with ${\rm
  sin}^2 \theta_W^{\ell\ {\rm eff}} = 0.23113\ (21)$ and
$CL(1.6/2)=0.44$. In contrast, the three hadronic measurements share
challenging experimental and theoretical systematic issues, including
heavy flavor tagging, large QCD corrections, and, especially, reliance
on hadronic Monte Carlo simulations to merge the QCD corrections with
the experimental acceptance. They combine to give ${\rm sin}^2
\theta_W^{\ell\ {\rm eff}} = 0.23222\ (27)$ with $CL(0.02/2)=0.99$.
The surprisingly small \chisqsp results from an underlying 14
parameter heavy flavor fit, with an even more surprising \chisq,
$CL(53,91)= 0.9995$. These small \chisqsp values could result from
overestimated systematic errors, but then the significance of the
discrepancy is exacerbated and the fit CL decreases: e.g., using just
statistical errors for the three hadronic measurements the \chisqsp of
the SM fit increases to 20.2/12 and the CL falls to 0.06. Another
possible explanation of the small \chisqsp values is that they reflect
incompletely understood correlations, which would again point to the
possiblity of underestimated systematic error. A more detailed
discussion is given in  the talk cited in \cite{msc123}.

Only future experimental results can help us to choose among the three
possible explanations. In this work we focus on the third possibility,
not because we know it to be more likely but rather to understand the
consequences. It might appear at first glance that the problem for the SM
would be resolved if the three hadronic asymmetry measurements are
assumed to have underestimated systematic errors and are omitted from
the fit.  The \chisqsp fit is then robust, with the p-value increasing
from $CL(17.2,12)=0.14$ to $CL(5.63,9)=0.78$, but the prediction for
\mhsp becomes problematic, with central value $m_H= 50$ GeV, with the
95\% CL upper limit at $m_H= 105$ GeV, and with only 3\% probability
in the region allowed by the LEP II lower bound, $CL(m_H >
114)=0.031$, excluding the SM at $\simeq 97\% CL$. The \mhsp
predictions from the two fits are summarized in table 3.

\begin{table}
\begin{center}
\vskip 12pt
\begin{tabular}{c|ccc}
\hline
\hline
 & $m_H$ (GeV) & 90\% $CL$ & $CL(m_H\, >\, 114)$ \\ 
\hline
\hline
All data & 85 & $ 47 < m_H < 158 $ & 0.26 \\
\hline
$A_{FB}^b \oplus A_{FB}^c \oplus Q_{FB}$ excluded& 50 & $24 <m_H< 105$ & 0.03\\
\hline
\hline
\end{tabular}
\end{center}
\caption{Predictions for $m_H$ from fits with and without the hadronic
asymmetries.}
\end{table}

One might think since the \chisqsp CL and $CL(m_H > 114)$
are independent probabilities that their product would be a measure 
of the likihood that the data agrees with the SM both as to 
the precision measurements and the Higgs mass prediction. However the 
product is not a fair estimator because it does not reflect the many ways that 
two independent probabilities can yield a product of  
a given value $P_1P_2$. A better estimator is the combined probability 
$P_C$ that the product of two independent,
uniformly distributed probabilities is less than or equal to the
product $P_1 P_2$, which it is easy to show is given by 
$$
P_C(P_1P_2)= P_1P_2 (1 - {\rm log}(P_1P_2)).       \eqno{(2)}
$$ 
The fit to all data then yields $P_C(0.14\cdot 0.26)= 0.16$, little
changed from the \chisqsp likelihood alone, while the reduced fit
yields a somewhat smaller value, $P_C(0.78\cdot 0.03)= 0.11$. Clearly
the consistency of the SM with the data is not improved by
removing the hadronic asymmetry measurements but rather the nature 
of the problem changes while remaining no less severe. 

Like the discrepancy between the hadronic and leptonic asymmetry
measurements, the conflict of the reduced data set with the LEP II
bound also has the canonical three possible generic explanations: new
physics, systematic error or statistical fluctuation. We focus here on
the possibility that it is an indicator for new physics, and consider
below a class of \zpsp models that can maintain the quality of the
\chisqsp fit for the reduced data set while raising the \mhsp
prediction into the allowed region above 114 GeV.

\noindent {\bf 3. \zpsp Models} 

We follow the framework described in \cite{ceg,adh} and explored in
detail in \cite{adh}. Restricting the fermionic content to the
three SM generations augmented just by three right-handed neutrinos
and assuming that the \zpsp couples universally to the three
generations, a new $U(1)_X$ gauge group is constrained to act on the
three fermion generations like an arbitrary linear combination of the
SM hypercharge $Y$ and $B-L$, the difference of baryon and lepton
number, as in equation (1).\cite{ceg,adh} Our study is restricted to
the case of a very heavy \zpsp boson. Referring to the original,
unmixed heavy gauge boson as \zzerop, we assume that the \zzeropsp
mass is generated primarily by a heavy SM-singlet Higgs boson \hpsp
with a large vacuum expectation value, $v^{\prime} \gg v = 246$ GeV,
and that $m_{Z^{0\, \prime}} \gg m_Z^0$. In order to preserve the
$U(1)_X$ gauge invariance of the SM Yukawa interactions, $Q_X$ must
also act on the SM Higgs boson $H$ as indicated by equation (1).  The
interaction of \zzeropsp with $H$ then gives rise to mass mixing
between \zzeropsp and the SM boson $Z^0$, resulting in the mass
eigenstates $Z$ and \zp. In this framework $Z^0-Z^{0\, \prime}$ mixing
only occurs if $\theta_X \neq 0$.

The upper $2\times 2$ corner of the $3 \times 3$ $W_3 - B - Z^{0\,
  \prime}$ mass matrix can be block diagonalized, yielding the
massless photon eigenstate and the residual $2\times 2$ $Z^0 - Z^{0\,
  \prime}$ mass matrix, written compactly as
$$
{\cal M}^2= m_{Z^0}^2\left( \begin{array}{cc}
                              1 & -r \cos\theta_X \\
                            -r \cos\theta_X & \hat{m}_{Z^{0\, \prime}}^2
                            \end{array}   \right).    \eqno{(3)}  
$$             
In equation (3) $m_{Z^0} = g_Z v/2$ is the usual unmixed $Z^0$ boson
mass, where $g_Z = g/\cos \theta_W$, $g$ is the $SU(2)_L$ gauge
coupling constant, and $\theta_W$ is the weak interaction mixing
angle.  The quantity $r$ is the ratio of the $U(1)_X$ gauge coupling
$g_{Z^{\prime}}$ to $g_Z$,
$$
r = {g_{Z^{\prime}} \over g_Z}    \eqno{(4)}
$$
and $\hat{m}_{Z^{\prime}}$ is the ratio of the \zpsp mass to the $Z$ mass, 
$$
\hat{m}_{Z^{\prime}}= {m_{Z^{\prime}} \over m_Z} 
           \simeq {m_{Z^{0\, \prime}} \over m_Z^0}\gg 1.    \eqno{(5)}
$$
Diagonalizing the mass matrix the leading correction to 
the $Z$ boson mass is 
$$
\delta m_Z^2 = -r^2 \cos^2\theta_X\, {m_Z^2\over \hat{m}_{Z^{\prime}}^2}   \eqno{(6)}
$$
and the $Z - Z^{\prime}$ mixing angle $\theta_M$, defined by 
$$
Z= \cos\theta_M \, Z^0 +\sin\theta_M \, Z^{0\, \prime}     \eqno{(7a)}
$$
$$
Z^{\prime}= \cos\theta_M \, Z^{0\, \prime} +\sin\theta_M \, Z^0,   \eqno{(7b)}
$$
is
$$
\theta_M= {r \cos\theta_X \over \hat{m}_{Z^{\prime}}^2}.  \eqno{(8)}
$$ 
Per equation (5), equations (6) and (8) are  correct to leading 
order in $1/\hat{m}_{Z^{\prime}}^2$.

The effect of the shift in the $Z$ boson mass on the radiative corrections 
can be encoded\cite{bh} as a contribution to the oblique parameter
$T$,\cite{pt}
$$
\alpha T_X= -{\delta m_Z^2 \over m_Z^2}  \eqno{(9)}
$$
so that
$$
\alpha T_X=  {r^2 \cos^2\theta_X \over \hat{m}_{Z^{\prime}}^2}. \eqno{(10)}
$$
The negative sign in equation (6), that occurs 
because the ``levels repel'' in two body mixing, implies a positive 
sign for $T_X$, which then causes the EW fit to prefer larger values 
of the Higgs boson mass.  

The second manifestation of $Z-Z^{\prime}$ mixing on the radiative
corrections is the shift in the $Z \overline ff$ couplings due to the
admixture of \zzeropsp in the $Z$ mass eigenstate. Including the oblique
corrections the interaction is
$$
{\cal L}_f= g_Z \left(1 + {\alpha T_X \over 2}\right) g_f^{\prime}
            \overline f {\not\! Z} f
	                         \eqno{(11)}
$$
where $f$ represents a quark or lepton of chirality $L$ or $R$
and $g_f^{\prime}$ encodes the $Z \overline ff$ coupling,
$$
g_f^{\prime}=  g_f + r\theta_M q_X^f 
	                         \eqno{(12)}
$$
Here $g_f$ is the SM $Z \overline ff$ coupling
$$
g_f= t_{3L}^f -q^f \hat{x}_W
	                         \eqno{(13)}
$$
where $t_{3L}^f$ and $q^f$ are the weak isospin and electric charge of
fermion $f$. The quantity $\hat{x}_W$ in equation (13) is the
oblique-corrected square of the $\sin$ of the SM weak mixing angle, $x_W=
\sin^2\theta_W$,
$$
 \hat{x}_W- x_W= -{x_W(1-x_W)\over 1-2x_W}\alpha T_X, 
                                 \eqno{(14)}
$$
and $q_X^f$ is the $Q_X$ charge of fermion $f$,
$$
q_X^f = \cos\theta_X\ {y^f \over 2} 
           +\sin\theta_X\ { b^f -  l^f\over 2}
                         \eqno{(15)}
$$
where $y^f,b^f, l^f$ are respectively the weak 
hypercharge, baryon number, and lepton number of fermion $f$.
In keeping with the approximation $\hat{m}_{Z^{\prime}} \gg 1$ we kept only 
the leading term in $\theta_M$ in equation (12).

For a given choice of $\theta_X$ the effect of $Z-Z^{\prime}$ mixing 
on the EW fit is determined by a single parameter, which we  choose to be 
$T_X$. The shift in the $Z\overline ff$ coupling, equation (12) is 
determined by 
$$
\epsilon= r \, \theta_M,
             \eqno{(16)}
$$
which, using equations (8) and (10) is determined by \tx,
$$
  \epsilon= {\alpha T_X \over  \cos \theta_X}.   \eqno{(17)}
$$
The \chisqsp fits presented in the section 5 are obtained by
scanning over $T_X$ in addition to the four SM scanning
parameters, $m_t, \Delta \alpha^{(5)}_{\rm had}(m_Z),\alpha_S(m_Z)$, and
$m_H$. The value of $T_X$ determines the ``effective Fermi 
constant'' of the \zpsp boson, defined as 
$$
G_{Z^{\prime}} = {g_{Z^{\prime}}^2 \over 4\sqrt{2}\, m_{Z^{\prime}}^2}. 
               \eqno{(18)}
$$
Defining $G_Z$ analogously,
$$
G_Z = {g_Z^2 \over 4\sqrt{2}m_Z^2}, 
               \eqno{(19)}
$$
which is equal at leading order to the Fermi constant, $G_Z=G_F$,
we have 
$$
 \hat{G}_{Z^{\prime}} = {G_{Z^{\prime}} \over G_Z} 
          = {r^2\over \hat{m}_{Z^{\prime}}^2}
          = {\alpha T_X\over \cos^2 \theta_X}.
                                 \eqno{(20)}
$$
Since $G_{Z^{\prime}}$ is constrained by the LEP II bounds, for a 
given value of $\theta_X$ we obtain constraints on $T_X$ both 
from the EW fits and from the LEP II bounds.

Before proceeding to the EW fits of the \zpsp models we briefly
mention an amusing feature of the $Y$-sequential models. In general
the EW corrections from \zzpsp mixing include both an oblique
correction \txsp from the shift in the $Z$ boson mass\cite{bh} and
non-oblique corrections from shifts in the $Z\overline ff$ couplings,
equation (12), due to the \zzeropsp component of the $Z$ eigenstate. But
for the case of a $Y$-sequential \zpsp boson, $\theta_X=0$, we find
that both the oblique and non-oblique corrections can be fully
parameterized by correlated ``pseudo-oblique'' parameters, \spsp and
\tp, defined by
$$
T^{\prime}= -T_X           \eqno{(21a)}
$$
$$
S^{\prime}= -4(1-x_W)T_X           \eqno{(21b)}
$$
where now 
$$
\alpha T_X = \epsilon= -{\delta m_Z^2 \over m_Z^2}. \eqno{(22)}
$$
The precision EW fit for the $Y$-sequential \zpsp
boson model can then be extracted from the usual oblique fit by
considering the line $S= 4(1-x_W)T$ in the $S,T$ plane with $T < 0$. 

This parameterization of the model immediately reveals that the value
of \mhsp cannot be increased toward the TeV scale and into the domain
of dynamical symmetry breaking, which requires positive $T$ and small
or negative $S$.\footnote{See for instance figures (12) and (13) of
  the second paper cited in \cite{msc123}.}  Since the original model
with $T_X > 0$ yields the same physics as the pseudo-oblique
representation with $T^{\prime} < 0$, because of the compensating
effects of $\epsilon$ and \sp, we also see that one cannot attach an
absolute significance to the sign of weak isospin breaking.

The equivalence of the two representations is explained by the fact
that for the $Y$-sequential model the apparently non-oblique
correction to the $Z\overline ff$ couplings induces a 
rescaling of the SM hypercharge coupling constant, which in turn 
contributes to $W_3$-$B$ kinetic mixing parameterized by $S$.
This is not true for the other models we consider with $\theta_X
\neq 0$, since the term proportional to $B - L$ cannot be absorbed
into a renormalization of any SM interaction.
A derivation is presented in the Appendix.

\noindent{\bf 4. Direct limits on \zpsp bosons from LEP II and CDF}

Carena {\it et al.}\cite{cddt} have used LEP II bounds\cite{lep2} on
contact interactions to extract limits on a variety of \zpsp bosons.
Their results constrain the \zpsp effective Fermi constant, that is,
the ratio of \zpsp mass to coupling strength, and in some  
cases they provide a stronger constraint than the precision EW data.
For the interesting class of models with $0 \leq \theta_X < \pi/2$ the
CDF collaboration has obtained bounds\cite{cdf} which are stronger
that the LEP II bounds if $g_{Z^\prime}$ is sufficiently small, $g_{Z^\prime}
\ \ltap\ g_Z/4$.  Both direct and EW constraints are presented in the
results presented below. In this section we summarize the LEP II and
CDF constraints for the \zpsp bosons considered in the EW fits
presented in section 5 below.

The class of Abelian charges considered here, defined in equation (1),
is equivalent, in the notation of Carena {\em et al.} to the group
$U(1)_{q+xu}$, characterized by the parameter $x$ which ranges from
$-\infty$ to $+\infty$ --- see their table I. It is easy to see that
the corresponding charge is
$$
\hat Q_X= {x-1 \over 3}Y + {4-x\over 3}(B-L)     \eqno{(23)}
$$
so that their $x$ is related to our $\theta_X$ by
$$
\tan \theta_X = {4-x\over x-1}.     \eqno{(24)}
$$
Defining $\hat g_{Z^{\prime}}$ as the corresponding coupling constant,
the relation between the coupling constants, determined 
by $\hat g_{Z^{\prime}} \hat Q_X= g_{Z^{\prime}} Q_X$, is 
$$
g_{Z^{\prime}} = {2 \over 3}\hat g_{Z^{\prime}} 
            \sqrt{2x^2 -10x +17}.   \eqno{(25)}
$$

With this dictionary we can translate the bounds obtained in
\cite{cddt} to the notation used here.  We will see below that the
most interesting region in $\theta_X$ for the precision fits is the
first quadrant, $0 \leq \theta_X < \pi/2$, corresponding to the
interval $4 \geq x > 1$. Within this interval the 95\% CL limit is
(see figure 1 of \cite{cddt})
$$
{m_{Z^{\prime}} \over \hat g_{Z^{\prime}}} > (2.62 + 1.18x){\rm TeV}
                              \eqno{(26)}
$$
Using the dictionary, equations (24) and (25), and equations 
(4), (5), and (10), equation (26) implies a bound on $T_X$,
$$
T_X \leq {\alpha^{-1}(m_Z) \over (30.1 + 15.5 \tan \theta_X)^2},
              \eqno{(27)}
$$ 
valid for $0 < \theta_X \leq \pi/2$. However, it should be noted that
for $\theta_X$ very near $\pi/2$ there must be a stronger bound, as can be
seen by comparing the limits on $U(1)_{q+xu}$ and $U(1)_{B-xL}$ in
figure 1 of \cite{cddt}. At $x=1$ both of these $U(1)$'s become $B-L$
but in the figure the latter is bounded more strongly than the former.
A stronger bound must then exist on \zpsp bosons with charge \qxsp for
$\theta_X$ near $\pi/2$, that could be extracted from the two lepton
differential cross sections, which are however not publicly available.
In the following we restrict ourselves to the conservative bound,
equation (27).\footnote{I thank the authors of \cite{cddt} for
  correspondence on this point.}

Although the models with the greatest effect on \mhsp lie in the first
quadrant, $0 \leq \theta_X < \pi/2$, it is also interesting to
consider the case $Q_X = T_{3R}$, since it occurs in attractive
left-right extensions of the SM and also because it is typical of
models in the second quadrant (or, equivalently, the fourth quadrant,
since only the sign of $g_{Z^{\prime}}\cdot Q_X$ is physical). For
$Q_X = T_{3R}$ we have $\theta_X = -\pi/4$, which corresponds to $x
\to \infty$ in the notation of \cite{cddt}. The bound for this case is
not discussed in \cite{cddt}, and we have extracted it directly from
the LEP II constraint on the $RR$ contact interaction quoted in
\cite{lep2}. In addition to $T_{3R}$ we will sample the following
choices from the first quadrant: $\theta_X = 0, \pi/6, \pi/3$, and
$11\pi/24$, for which the LEP II bounds on $T_X$ and \hatgzpsp are
given in table 4.

The case of $\theta_X = 11\pi/24$ is interesting because we will see
in the next section that it has an appreciable effect on the EW fit
even though it is very near $\theta_X = \pi/2$ corresponding to $Q_X=
B-L$, for which there is no \zzpsp mixing and therefore no effect on
the EW fit.  However the surprisingly large effect that is found on
the EW fit is severely constrained by the direct limit from equation
(27) quoted in table 4.

\begin{table}
\begin{center}
\vskip 12pt
\begin{tabular}{c|cc}
\hline
\hline
 $\theta_X$ & $T_X$ & $\hat G_{Z^{\prime}}$ \\
\hline
\hline
0 &0.14 & 0.0011\\
$\pi/6$ &0.084 &0.00088 \\
$\pi/3$ &0.039 & 0.0012\\   
$11\pi/24$ &0.0059 & 0.0027\\
$-\pi/4$ &0.30 & 0.0047\\
\hline
\hline
\end{tabular}
\end{center}
\caption{95\% CL upper limits on $T_X$ and $\hat G_{Z^{\prime}}$ obtained 
from LEP II bounds on contact interactions.}
\end{table}

\begin{table}
\begin{center}
\vskip 12pt
\begin{tabular}{c|ccc}
\hline
\hline
 $\theta_X$ &r & $m_{Z^\prime}$ (TeV) &$T_X$  \\
\hline
\hline
  & 0.27  & 0.83  & 0.11  \\
 0& 0.13  & 0.70  & 0.039  \\
  & 0.081  &0.61   &0.019   \\
\hline
  & 0.20  & 0.78  & 0.051  \\
 $\pi/6$ &0.098   & 0.64  & 0.019  \\
  & 0.059  &0.54   &0.0095   \\
\hline
  &  0.20 &0.75   &0.018   \\
 $\pi/3$  &0.098   &0.60   &0.0070   \\
  &0.059   &0.45   & 0.0046  \\
\hline
  &0.24   &0.69   &0.0022   \\
$11\pi /24$  &0.12   &0.50   &0.0010   \\
  & 0.072  &0.40   &0.00058   \\
\hline
\hline
\end{tabular}
\end{center}
\caption{95\% CL upper limits on $T_X$ and $m_{Z^{\prime}}$ 
from CDF\cite{cdf} for given values of $r=g_{Z^\prime}/g_Z$.}
\end{table}

The corresponding bounds on \txsp from the CDF collaboration\cite{cdf}
are given in table 5, translated from the notation of reference
\cite{cddt}, which is followed in reference \cite{cdf}, to the
notation used here. For each $\theta_X$ in table 4, except $\theta_X =
-\pi/4$ for which no bound is given by CDF, we present the implied
limit on $m_{Z^\prime}$ and \txsp for given values of the \zpsp
coupling strength, parameterized as the ratio to the SM $Z$ boson
coupling, $r=g_{Z^\prime}/g_Z$. Comparing tables 4 and 5, we see that
the CDF bounds are stronger than the LEP II bounds for $r\ \ltap\
1/4$, becoming increasingly stronger as $r$ decreases.  The LEP II
bounds depend only on the ratio $g_{Z^\prime}/m_{Z^\prime}$, i.e., on
the effective Fermi constant \gzp, independent of the value of
$g_{Z^\prime}$, because with $m_{Z^\prime}\gg m_Z$ they arise purely
from the interference of the high energy tail of the $Z$ boson
amplitude with the low energy tail of the \zpsp amplitude. At Fermilab
for sufficiently small $g_{Z^\prime}$, which corresponds to smaller
$m_{Z^\prime}$ for fixed \gzp, the data begins to be sensitive to the
direct production term, i.e., the square of the \zpsp amplitude,
giving rise to increased sensitivity and a stronger constraint.

\noindent {\bf 5. Electroweak Fits in \zpsp Models}

In this section we present fits to the precision EW data for the class
of \zpsp models discussed in section 3, focusing on the effect of
\zzpsp mixing on the value of the Higgs boson mass obtained from the
fits. We will use two statistical methods that illuminate the physics
in different ways, because they answer questions that are different in
detail though they are clearly related.  The first is the classical
frequentist method, which is used by the EWWG\cite{ewwg} and was used
in the discussion of the SM fit in section 2.  In this method the
question is `Without imposing any {\it a priori} knowledge, direct or
indirect, of the value of \mh, how well does the model describe the
precision data and what prediction does the best fit make for the
likelihood of different values of \mh?' The second approach, followed
for instance in the analysis of \zpsp models in \cite{flv}, might be
termed ``Bayesian,'' in the sense that it imposes external knowledge
about \mhsp as a ``prior'' constraint on the fit and therefore
assesses the extent to which the fit of the model to the precision EW
data is consistent with that prior. This approach then answers a
different question which might be stated as follows: `If the value of
\mhsp were known to have a specific value or to lie within a certain
range, how well does the model fit the precision data?'

Both questions are valid and interesting, and it is useful to see what
each tells us about the compatibility of various values of \mhsp with
the different models. We will consider each in turn, combining the
constraints from the fits with the  direct LEP II bounds on \zpsp bosons 
in table 4.

\noindent {\it 5a. Frequentist Fits}

We first present the frequentist fits for the models listed in table
4.  These fits contain one more free parameter than the SM fits, which
may be chosen to be the effective Fermi constant $G_{Z^{\prime}}$ or
equivalently $T_X$ (see equation (20)).  In the frequentist approach
to the SM fits, the global \chisqsp minimum is determined by freely
varying $m_t, \Delta \alpha^{(5)}_{\rm had}(m_Z),\alpha_S(m_Z)$, and
$m_H$.  The 95\% CL upper limit on \mhsp is then determined by
minimizing \chisqsp as \mhsp is varied away from its value at the
global minimum, until the local \chisqsp minimum (i.e., for the given
value of \mh) has increased by $\Delta \chi^2 = 2.71$, corresponding
to the upper boundary of the symmetric 90\% confidence interval for
one degree of freedom, $CL(2.71,1)=0.90$.

To extend this approach to the \zpsp models, we vary \txsp in addition
to $m_t, \Delta \alpha^{(5)}_{\rm had}(m_Z)$, $\alpha_S(m_Z)$, and
$m_H$, to obtain the global \chisqsp minimum, reducing the number of
degrees of freedom by one relative to the SM fit, and then vary both
\txsp and \mhsp about the \chisqsp minimum.  The 90\% contour in the
$m_H - T_X$ plane is then defined by $\Delta \chi^2 = 4.61$
corresponding to $CL(4.61,2)=0.90$, and similarly the 95\% contour is
at $\Delta \chi^2 = 5.99$. The allowed regions are then further
constrained by the 95\% exclusion limits on \txsp from the LEP II
bounds on contact interactions, table 4, which are superimposed over
the contours from the precision data.

We consider two data sets. Data set B excludes the three
hadronic asymmetry measurements. Set A contains all of the
measurements in table 1 except the jet charge asymmetry, \qfb, which
we omit for simplicity. (To compute the correction to \qfbsp we would 
have to convolute the
mixing-induced shifts in the $Z\overline qq$ couplings with the
$\overline qq$ partial rates, and then unfold the result to
obtain the effective value of \sinthl.) 
\qfbsp has very little impact on the fit CL or on \mh: figure 2
shows that \afbbsp completely dominates \afbcsp and \qfbsp in the
\chisqsp distribution, since the combined distribution is practically 
indistinguishable from the distribution of \afbbsp alone.

\begin{figure}[t]
\centerline{\includegraphics[width=3.6in,angle=90]{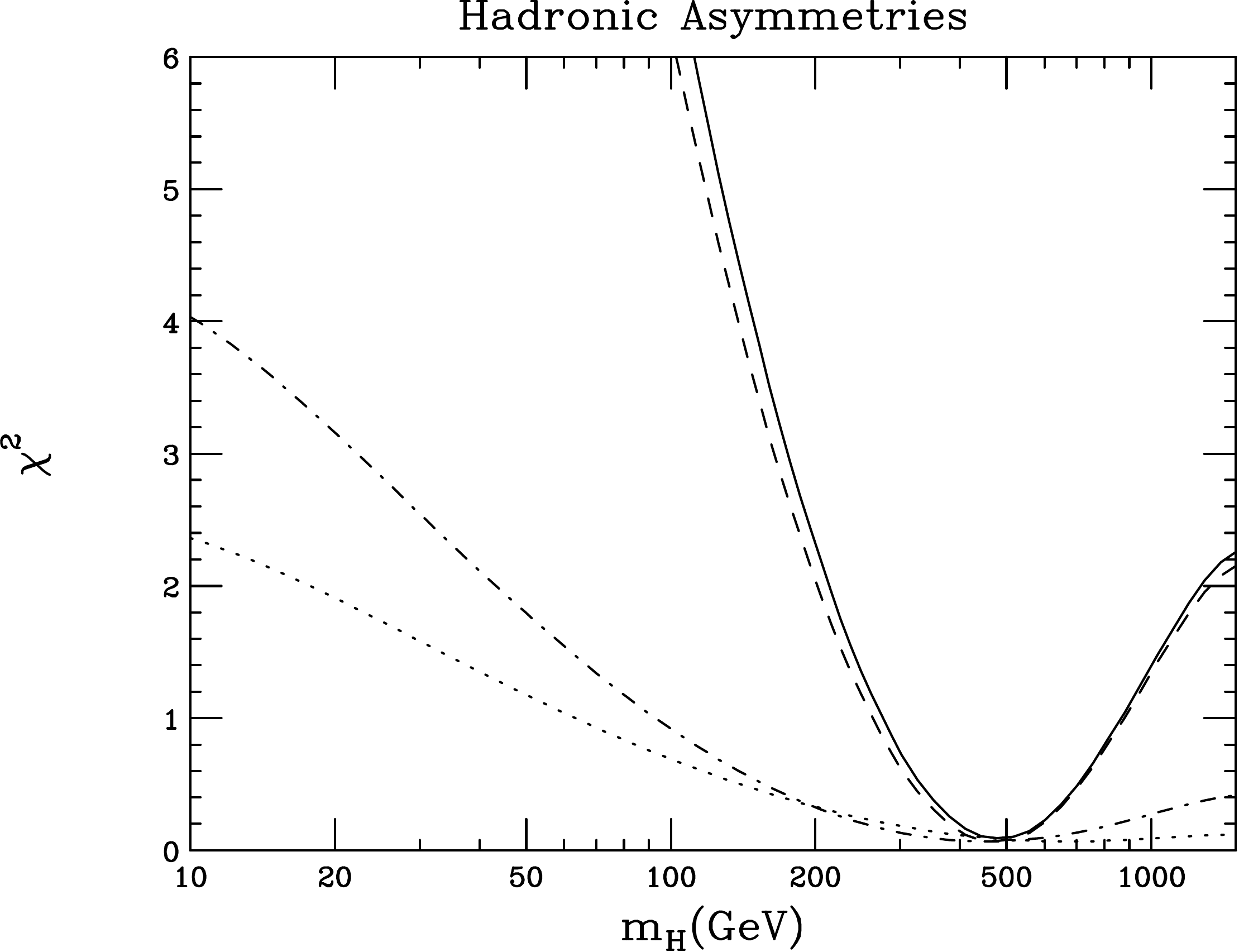}}
\caption{$\chi^2$ distributions as a function of $m_H$ for the 
combination of the three hadronic asymmetry measurements (solid line) 
and for each individually: \afbbsp (dashes), \afbcsp (dashdot), and 
\qfbsp (dots).}
\label{fig2}
\end{figure}

\begin{table}[t]
\begin{center}
\vskip 12pt
\begin{tabular}{c|ccccc}
\hline
\hline
 Data Set & $\chi^2/N$ & CL & $m_H$ (GeV) & $m_H(95\%)$ & CL($m_H > 114$) \\ 
\hline
\hline
{\bf A} & 16.5/11 & 0.12 & 85  & 153  & 0.24 \\
\hline
{\bf B} & 5.63/9 & 0.78 & 50  & 105  & 0.03 \\
\hline
\hline
\end{tabular}
\end{center}
\caption{SM fits A and B.  $m_H(95\%)$ is the usual frequentist 95\% upper 
  limit on \mhsp.}
\end{table}

\begin{table}[t]
\begin{center}
\vskip 12pt
\begin{tabular}{c||cccc||cccc}
\hline
\hline
    & \multicolumn{4}{c}{{\bf Data Set A}}& 
                \multicolumn{4}{c}{{\bf Data Set B}}\\
  $\theta_X$ & $\chi^2/N$ & CL & $m_H$ (GeV) & $T_X$ & 
             $\chi^2/N$ & CL & $m_H$ (GeV) & $T_X$ \\
\hline
0  & 16.5/10  &0.09  &85  &0.0 &5.56/8  &0.70  &61  &0.012  \\ 
 $\pi/6$ & 16.5/10  &0.09  &85  &0.0&5.39/8 &0.72  &70  &0.019  \\ 
  $\pi/3$ & 16.5/10  &0.09  &85  &0.0&5.22/8  &0.73  &70  &0.014  \\ 
  $11\pi/24$ & 16.5/10  &0.09  &85  &0.0&5.00/8  &0.76  &70  &0.006  \\ 
  $-\pi/4$ & 16.5/10  &0.09  &85  &0.0&5.63/8  &0.69  &50  &0.0  \\ 
\hline
\hline
\end{tabular}
\end{center}
\caption{Frequentist \chisqsp fits for \zpsp models.}
\end{table}

The SM fits to data sets A and B are summarized in table 6.
$m_H(95\%)$ is the frequentist 95\% CL upper limit, at $\chi^2 =
\chi^2_{\rm MIN} + 2.71$. Fit A has an acceptable prediction for \mhsp
but a marginal confidence level, while fit B has a robust \chisqsp CL
but a failed prediction for \mh.  The \zzpsp model fits to data sets A
and B are shown in table 7 and figures 3 - 7. The effect of \zzpsp
mixing on data set A is to push \afbbsp further from the experimental
value than the 2.82$\sigma$ deviation of the SM fit, so that the
minima for set A coincide with the SM and \chisqsp increases rapidly
away from the SM minimum. As shown in table 7, the \chisqsp minimum
for data set A is then at $T_X=0$ for all values of $\theta_X$,
implying zero mixing, $\theta_M = 0$.  The \chisqsp value and the
central value of \mhsp are then identical to the SM fit, while the
confidence level decreases since there is one fewer degree of freedom,
from CL(16.5,11) = 0.12 to CL(16.5,10) = 0.09. For set B the fits
favor nonzero but small mixing for models with $\theta_X$ in the first
quadrant, with modest decreases in the \chisqsp minimum and modest
increases in \mh, while the fit likelihoods decrease slightly. For
$T_{3R}$ as for all models with $\theta_X$ in the second quadrant, the
\chisqsp minimum is at zero \zzpsp mixing, except for $\theta_X$ very
near $\pi$ where $Q_X \simeq Y$.\footnote{Note that quadrants I and
  III in $\theta_X$ are physically equivalent, as are quadrants II and
  IV, since the overall phase of \qxsp is not physical because it can
  be compensated by the phase of $g_{Z^{\prime}}$.}

Although the changes in the \chisqsp minima are modest at best, there
is a substantial effect on the allowed range of Higgs boson masses in
the case of data set B and a smaller effect for data set A.  This can
be seen in the 90 and 95\% contours shown in figures 3 - 7.  For the
$Y$-sequential model, $\theta_X = 0$, in the case of data set B the
90\% contour extends to $m_H = 260$ GeV, a factor 2.5 beyond the 105
GeV 95\% upper limit of the SM fit.\footnote{The 90\% contour of the
  \zzpsp fit should be compared to the symmetric 90\% confidence
  interval of the SM fit, whose upper boundary defines the 95\% upper
  limit. The extreme of the 95\% contour corresponds to the 97.5\%
  upper limit of the SM fit.}  The LEP II upper limit from table 4,
$T_X < 0.14$, does not impinge on the contours from the EW fit. The
maximum reach in \mhsp occurs at $T_X= 0.10$, corresponding to $\hat
G_{Z^{\prime}}\simeq 8\cdot 10^{-4}$; if the \zpsp coupling were of
electroweak strength, this would imply a mass hierarchy of order
$m_{Z^{\prime}}/m_Z \simeq 30$ or $m_{Z^{\prime}} \simeq 3$ TeV,
within the range of the LHC. For fit A the increase in \mhsp is
smaller, with the extreme of the 90\% contour reaching 230 GeV, a
factor 1.5 above the SM 95\% upper limit at 153 GeV. Although the SM
fit of data set B predict smaller values of \mhsp than set A, with
\zzpsp mixing set B is consistent with larger values than set A.

\begin{figure} 
\centerline{\includegraphics[width=2.5in,angle=90]{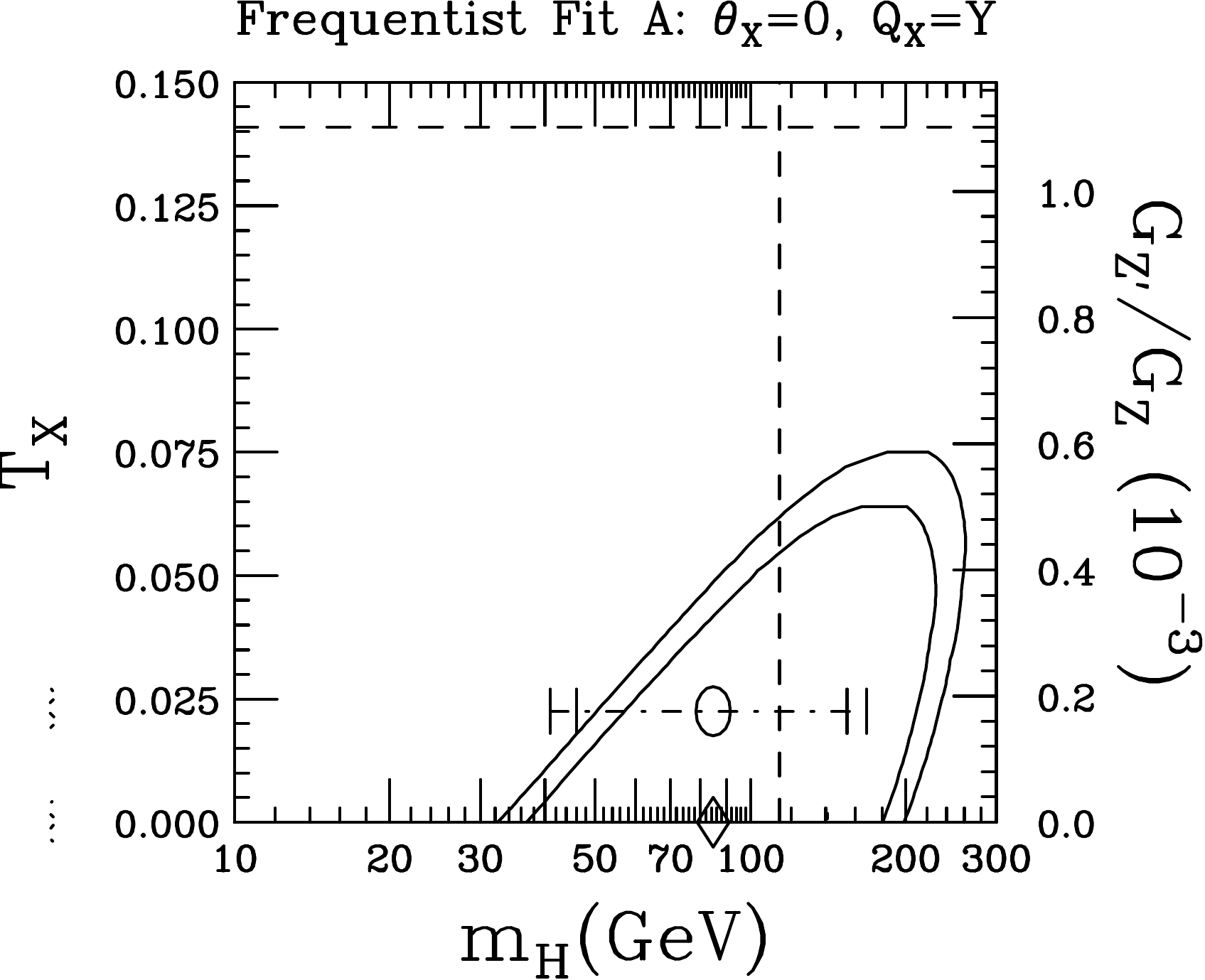}
\hspace*{.2cm}
\includegraphics[width=2.5in,angle=90]{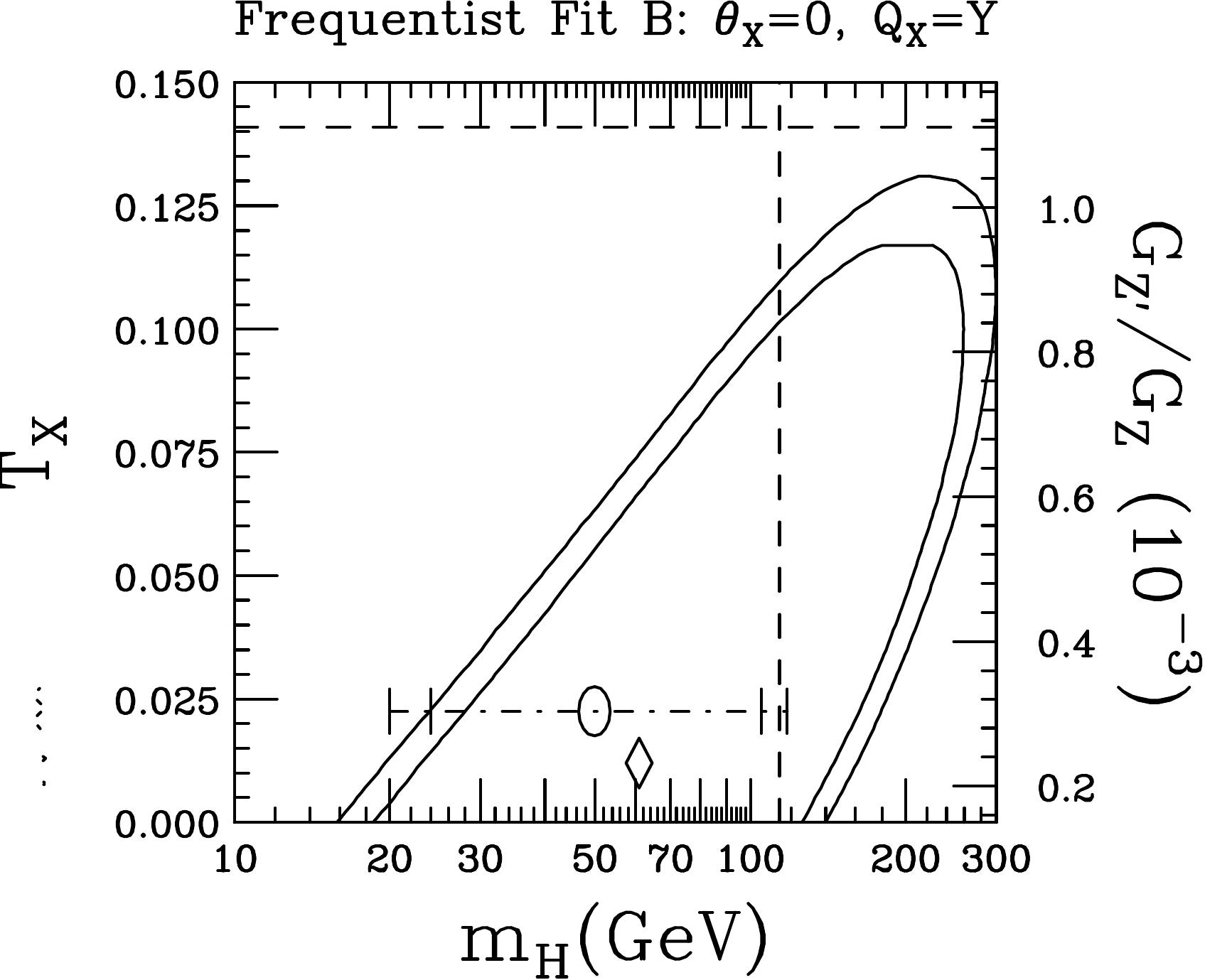}}
\caption{90\% and 95\% CL contours in the $T_X - m_H$ plane for
  frequentist fits to data sets A and B for the $Y$-sequential model,
  $\theta_X=0$. The right axis indicates the corresponding values of
  $\hat G_{Z^{\prime}}= G_{Z^{\prime}}/G_Z$ per equation (20). The
  diamond indicates $m_H, T_X$ at the \chisqsp minimum for the \zpsp
  model. (Note that for set A for all $\theta_X$ the \chisqsp minimum
  is at $T_X =0$ and the diamond is hiding on the x-axis.)  The ellipse and
  dot-dash horizontal line display the central value and 90, 95\%
  symmetric confidence intervals of \mhsp for the SM fit (elevated
  above $T_X = 0$ only for clarity). The horizontal dashed line is the
  95\% CL upper limit on \hatgzpsp extracted from LEP II data. The
  vertical dashed line is the LEP II 95\% lower limit on \mh.}
\label{fig3}
\end{figure}

\begin{figure}
\centerline{\includegraphics[width=2.5in,angle=90]{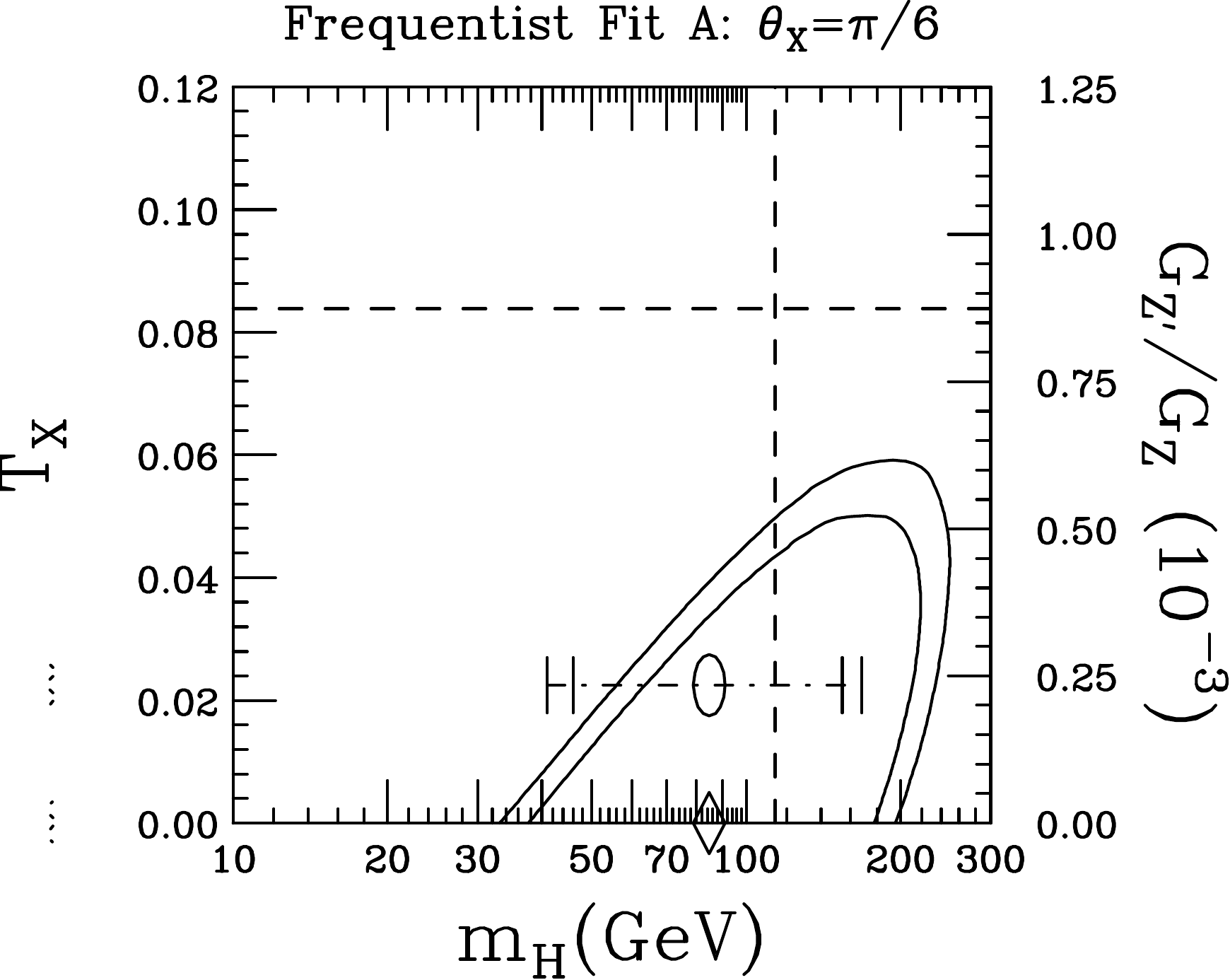}
\hspace*{.2cm}
\includegraphics[width=2.5in,angle=90]{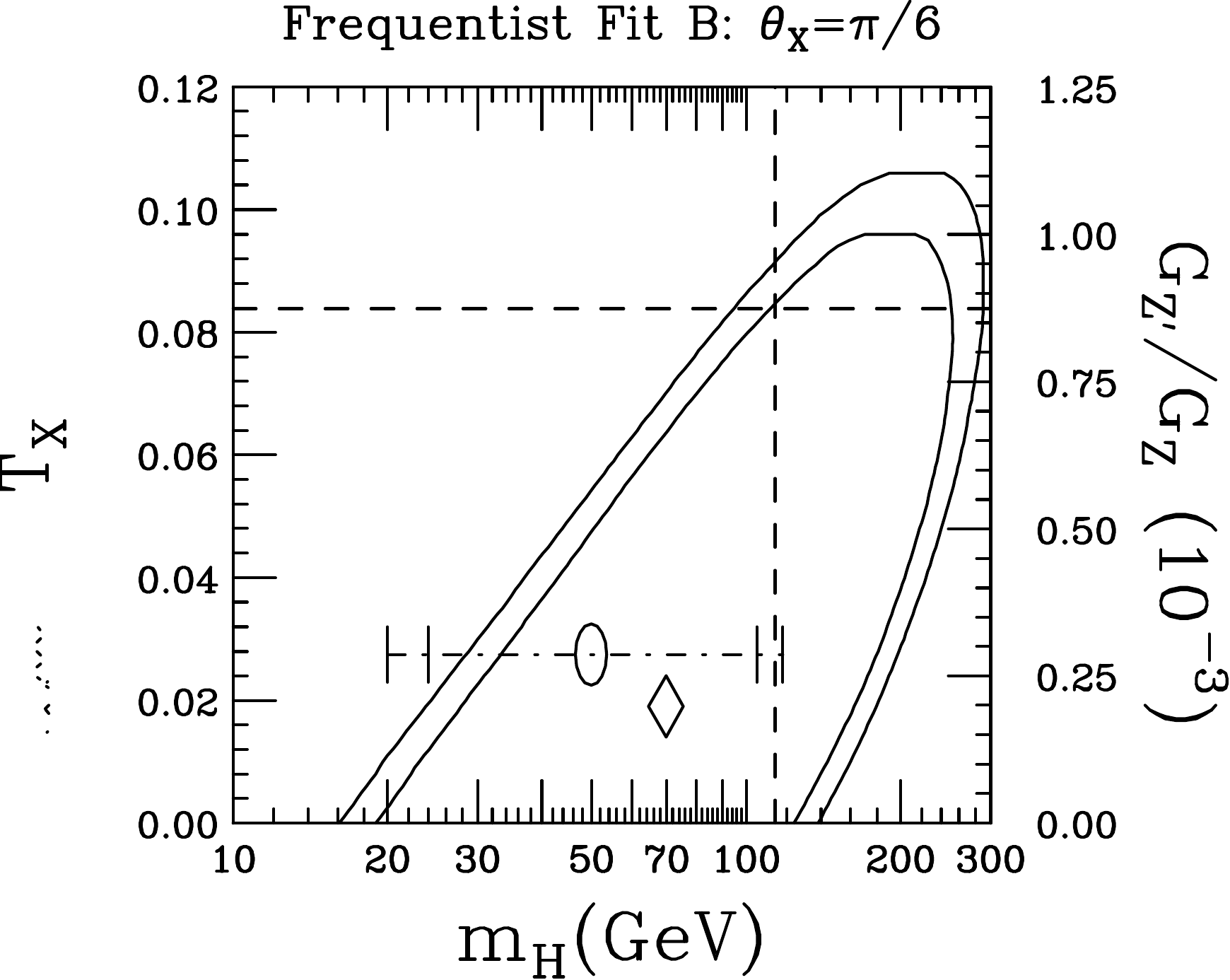}}
\caption{90\% and 95\% CL frequentist contours for \zpsp model with 
$\theta_X=\pi/6$, as in figure 3.}
\label{fig4}
\end{figure}

Figures 4 and 5 show that these features persist for $\theta_X =
\pi/6$ and $\pi/3$. The LEP II constraint on \gzpsp 
begins to limit the allowed region for data set B, both because
the LEP II bound becomes stronger and also because as $\theta_X$
increases toward $\pi/2$ the ratio of \gzpsp to \txsp increases like
$1/\cos^2 \theta_X$, as seen in equation (20). Even though $\theta _X
= 11\pi/24$ is very near $\pi/2$, $Q_X = B - L$, for which there is no
mixing and no effect on the fits, there is still a significant effect
on the allowed region in \mhsp for data set B, shown in figure 6.
However, the allowed range is severely constricted by the direct limit
on \gzp, which, as discussed in section 4, is likely to be even
stronger than is shown in the figure.  For $Q_X = T_{3R}$ the effect of
\zzpsp mixing on \mhsp is weaker, as can be seen from the more
vertical slopes of the contour lines above 114 GeV in figure 7, but
there is no additional constraint from the LEP II upper limit on
\gzpsp which is $T_X > 0.30$.

To estimate the confidence levels for these fits to lie within the LEP II
allowed regions for \mhsp and \gzpsp we use a Bayesian likelihood
method that was developed in the second paper cited in \cite{msc123}
to compute ${\rm CL}(m_H > 114 {\rm GeV})$ for the SM fits. In that
approach two Bayesian priors were introduced to convert unnormalized
likelihood functions into normalized probability distributions from
which confidence intervals could be extracted. The first prior is that
\mhsp lies between 10 and 3000 GeV. The precise value of the limits is
not critical since there is negligible support above 1000 GeV or below
10 GeV. The second prior is that $\log m_H$ is the appropriate
measure, a natural assumption since the EW corrections depend
logarithmically on \mh.

This procedure was shown to be reasonable (or at least no more foolish
than the conventional procedure) by the fact that it provided
confidence intervals for \mhsp similar to those obtained from the
$\Delta \chi^2$ method, e.g., for data set B the result was CL$(m_H
> 114)=0.030$ from the Bayesian likelihood method versus 0.035 from
$\Delta \chi^2$ with the data of the time. We now find CL$(m_H >
114)=0.17$ from the SM fit to data set A compared to 0.24 from $\Delta
\chi^2$, and 0.018 compared to 0.031 for set B. There is no reason
that the two methods should agree precisely. An important difference
is that the $\Delta \chi^2$ method compares only the {\em best} fits
at different values of \mh, while the Bayesian likelihood method
samples the complete distribution of scanned parameters ($m_t, \Delta
\alpha_5, \alpha_S$) at each value of \mh.

\begin{figure}
\centerline{\includegraphics[width=2.5in,angle=90]{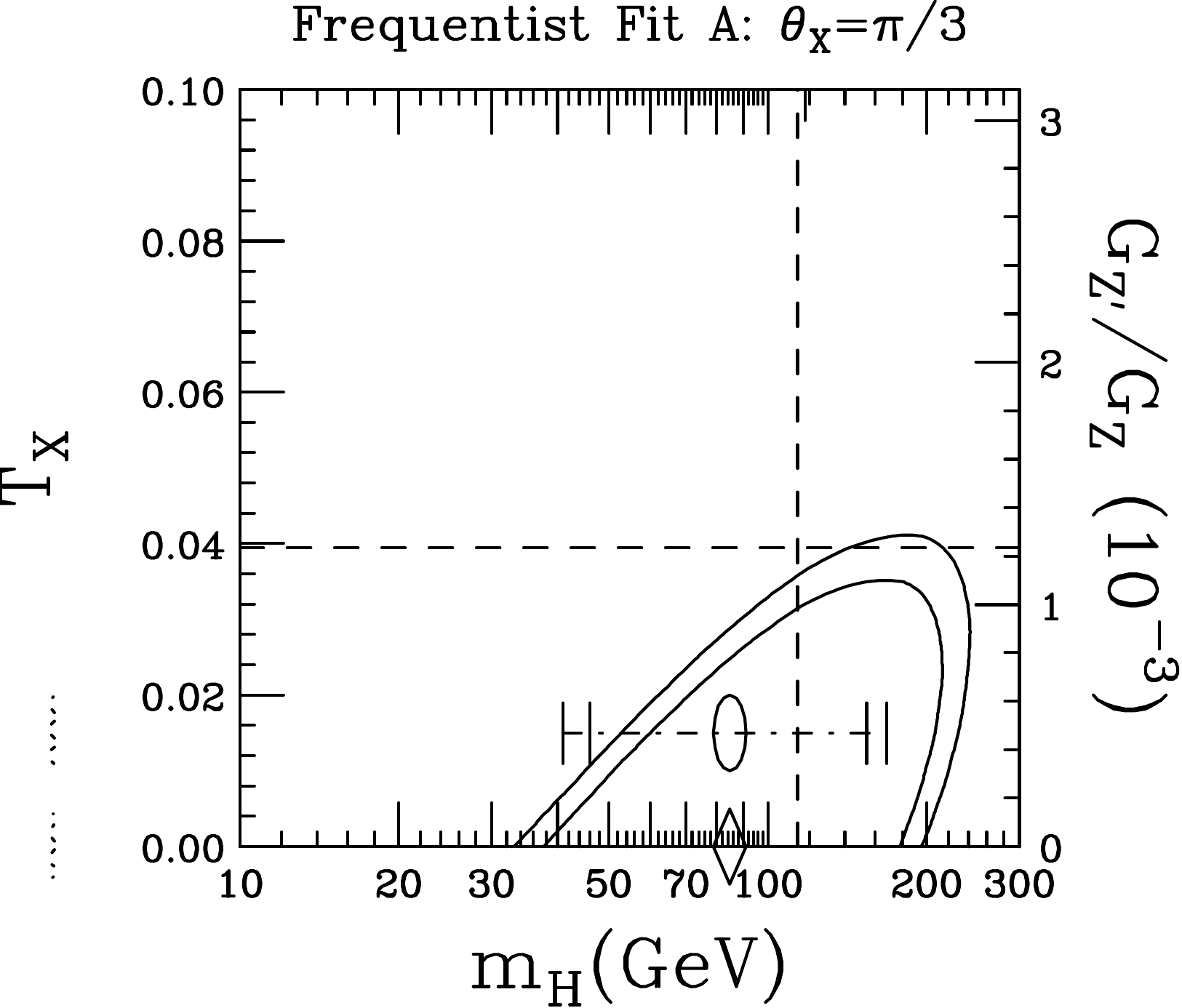}
\hspace*{.2cm}
\includegraphics[width=2.5in,angle=90]{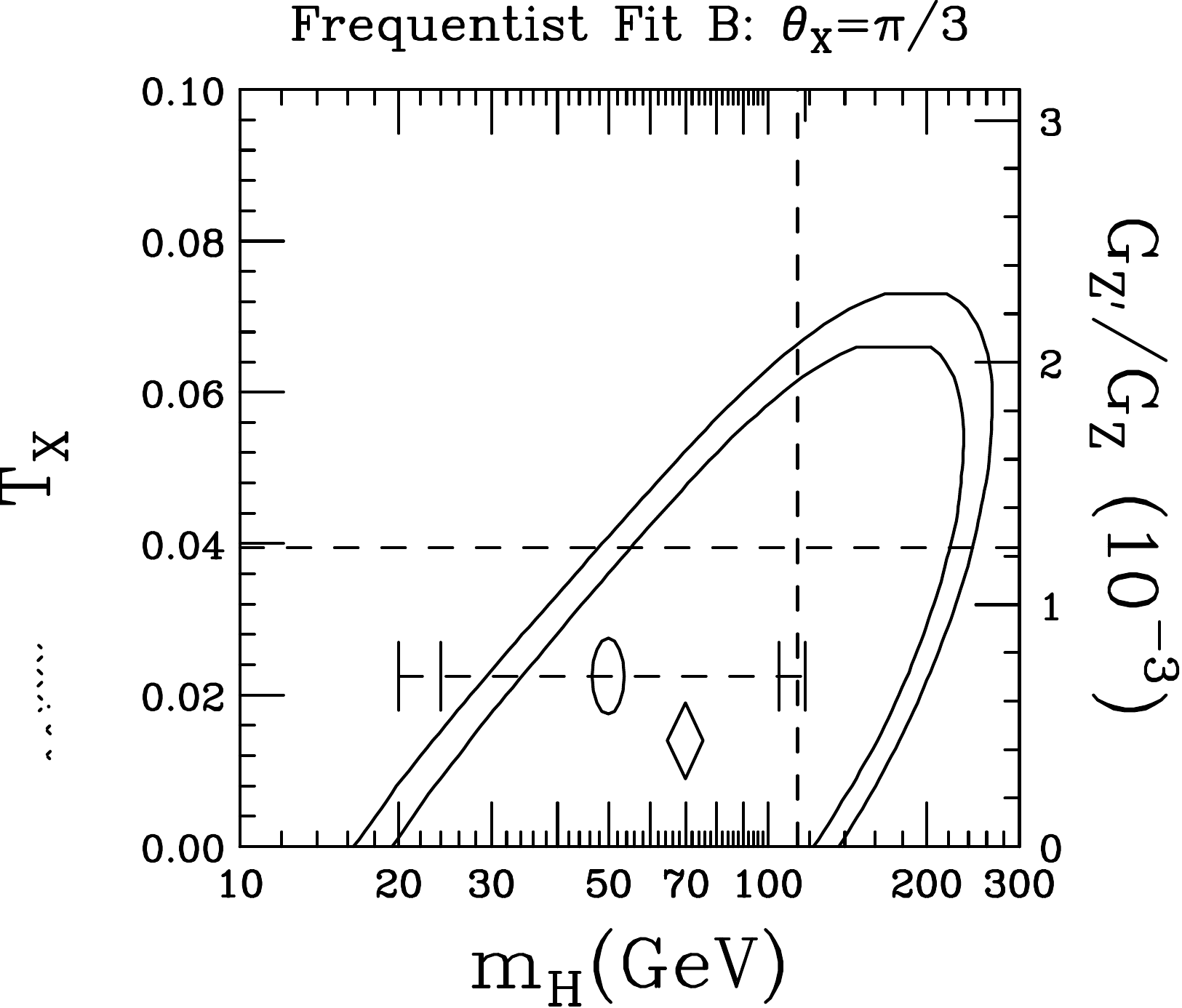}}
\caption{90\% and 95\% CL frequentist contours for \zpsp model with 
$\theta_X=\pi/3$, as in figure 3. }
\label{fig5}
\end{figure}

\begin{figure}
\centerline{\includegraphics[width=2.5in,angle=90]{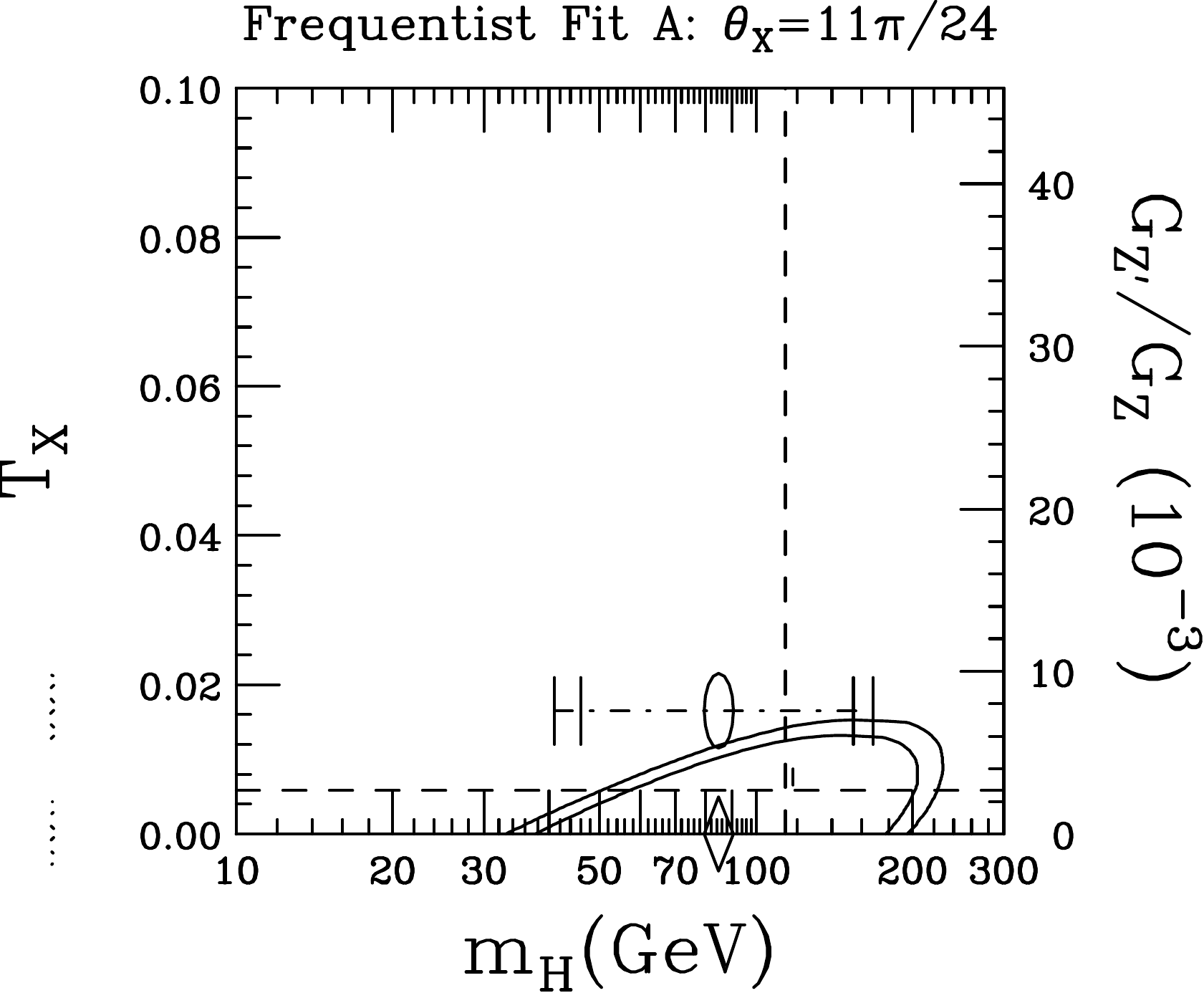}
\hspace*{.2cm}
\includegraphics[width=2.5in,angle=90]{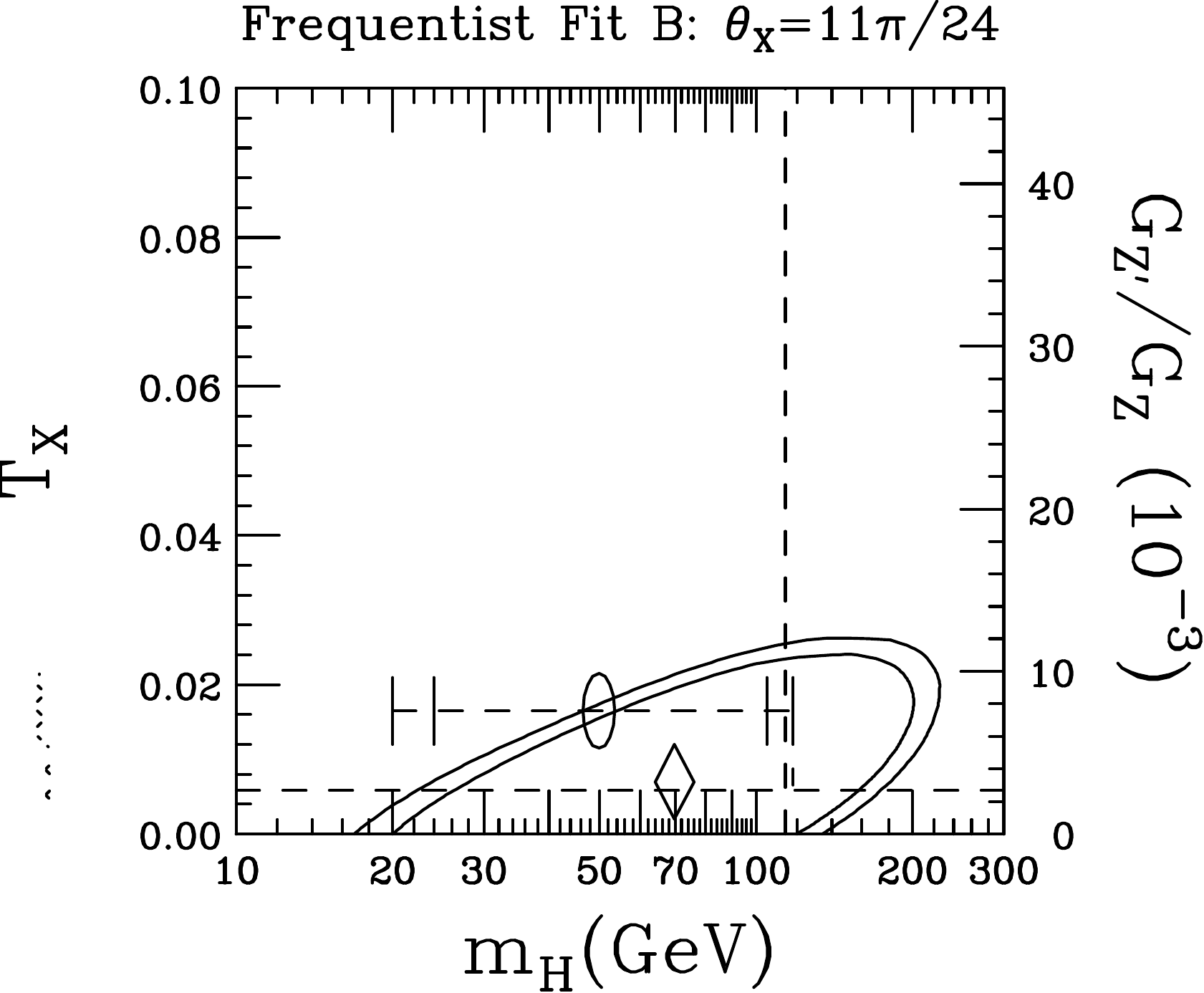}}
\caption{90\% and 95\% CL frequentist contours for \zpsp model with 
$\theta_X=11\pi/24$, as in figure 3.}
\label{fig6}
\end{figure}

\begin{figure}
\centerline{\includegraphics[width=2.5in,angle=90]{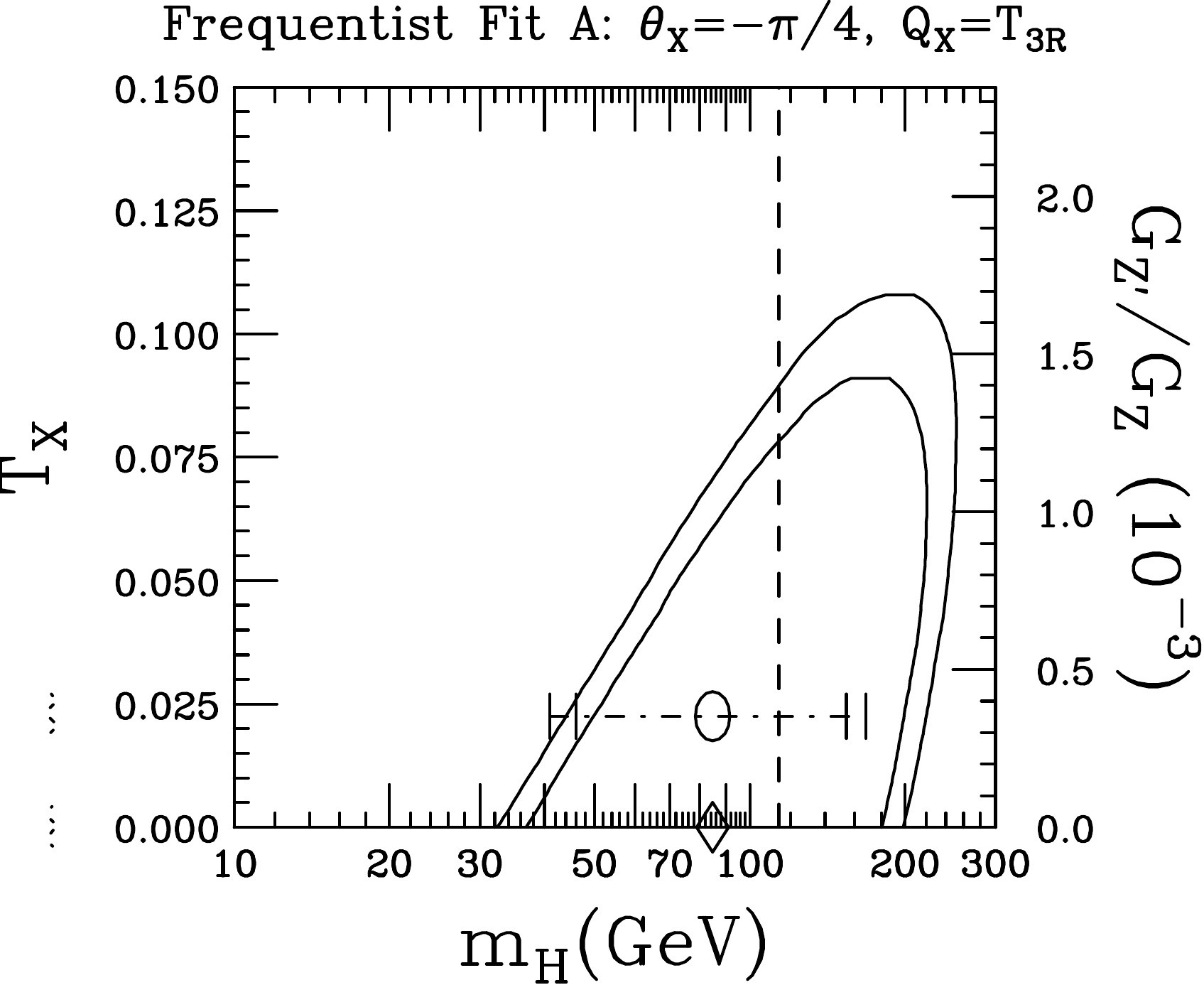}
\hspace*{.2cm}
\includegraphics[width=2.5in,angle=90]{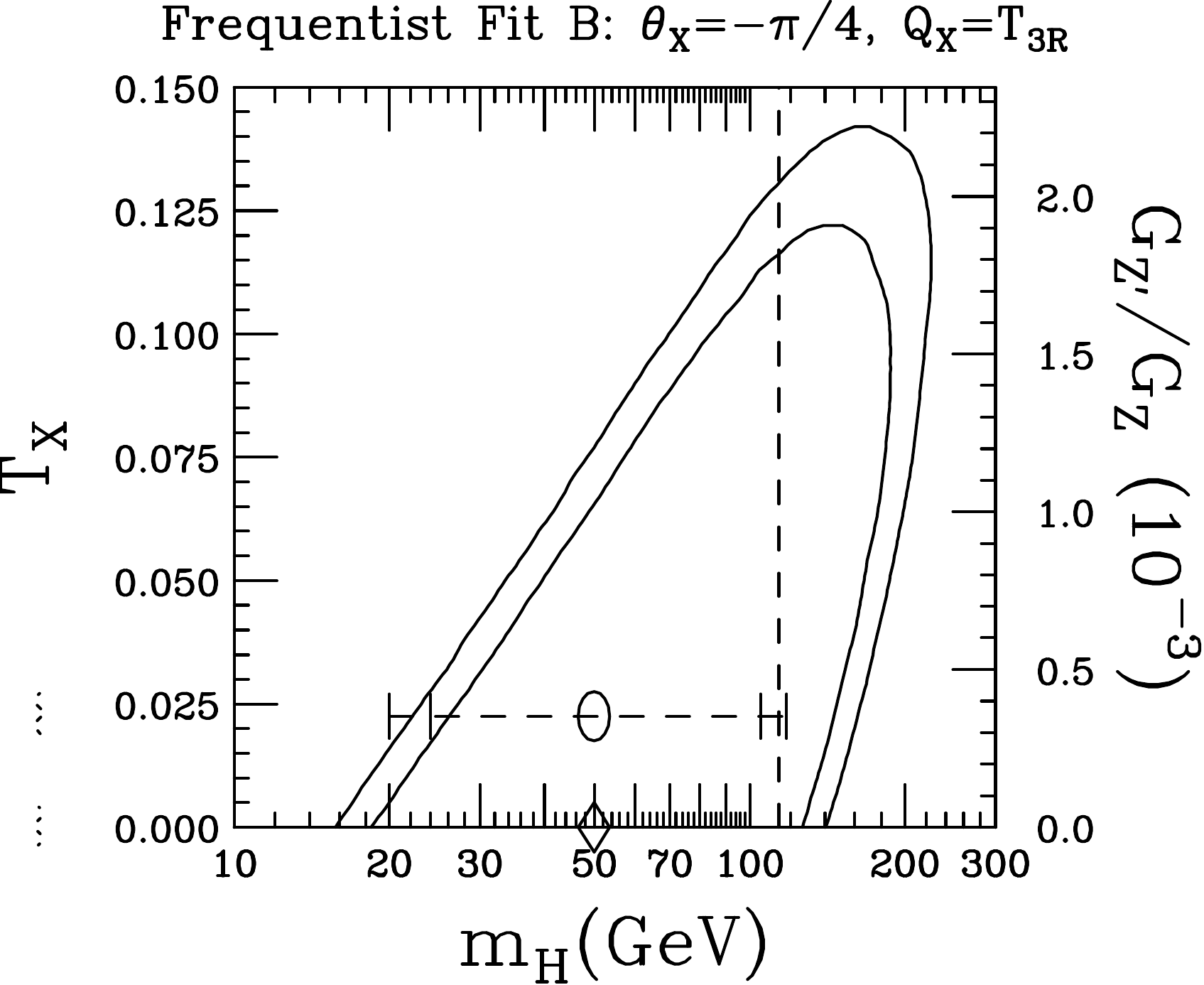}}
\caption{90\% and 95\% CL frequentist contours for \zpsp model with 
$\theta_X=-\pi/4$, $Q_X=T_{3R}$, as in figure 3. The LEP II upper limit, 
$T_X < 0.30$, is off the graph.}
\label{fig7}
\end{figure}

\begin{table}
\begin{center}
\vskip 12pt
\begin{tabular}{c||c||c|c|c}
\hline
\hline
 Data Set &Model &  $m_H(95\%)$ & CL($m_H> 114$) & 
               {\bf and} CL($T_X > T_{\rm LEP\ II}$) \\ 
\hline
\hline
{\bf A} & SM & 153 & 0.17 & ... \\
        &$\theta_X=0$ & 230 & 0.44 & 0.44 \\
        &$\pi/6$ &220 & 0.42 & 0.42 \\
        &$\pi/3$ &214 & 0.39 & 0.38 \\
        &$11\pi/24$ &202 & 0.34 & 0.15 \\
        &$-\pi/4$ &220 & 0.43 & 0.43 \\
\hline
{\bf B}  & SM & 105 & 0.018 & ... \\ 
        &$\theta_X=0$ & 260 & 0.29 & 0.29 \\
        &$\pi/6$ &252 & 0.28 & 0.24 \\
        &$\pi/3$ &221 & 0.23 & 0.12 \\
        &$11\pi/24$ &158 & 0.14 & 0.01 \\
        &$-\pi/4$ &188 & 0.18 & 0.18 \\
\hline
\hline
\end{tabular}
\end{center}
\caption{Fits of data sets A and B.  For the SM $m_H(95\%)$ is the 
  usual 95\% upper 
  limit obtained by the $\Delta \chi^2$ method. For the \zpsp models  
  $m_H(95\%)$ is the maximum value of \mhsp on the 90\% frequentist contours 
  (figures 3 - 7) that is 
  consistent with the 
  LEP II direct limit on \tx. 
  The confidence levels CL($m_H> 114$) and CL($T_X > T_{\rm LEP\ II}$) are 
  computed with the Bayesian likelihood method described in the text. 
  The entries in the last column 
  combine both the \mhsp and \txsp direct limits from LEP II.}
\end{table}

The same method can be applied to the two dimensional distributions in
\mhsp and \tx. The natural measure for \mhsp is again logarithmic.
Since \txsp represents a first order perturbation of new physics on
the leading order SM, the natural measure for \txsp is linear.  We
normalize the likelihood functions in the intervals $0 < T_X < 0.25$
and $10 < m_H < 3000$ GeV, where again the results are insensitive to
to the precise choice of limits.\footnote{$T_X>0$ is a boundary condition 
imposed by \zzpsp mixing.} The results are tabulated in table 8,
which displays the confidence levels for $m_H > 114$ GeV, both without
and with the LEP II constraint on \tx. In addition we tabulate
$m_H(95\%)$, which for the \zpsp models is defined as the largest
value of \mhsp on the 90\% contour that is consistent with the LEP II
bound on \gzp.  Unlike $m_H(95\%)$ for the SM fits, the values quoted
for the \zpsp models cannot be interpreted as reflecting a 5\% 
probability for $m_H > m_H(95\%)$. Again the impact
of \zzpsp mixing is greater for data set B, with the probability of
the LEP II allowed regions increasing by an order of magnitude relative 
to the SM value, e.g., from CL$(m_H\ >\ 114)=0.018$ for the SM to 
0.29 for the $Y$-sequential model. 

In table 9 we show the effect of the CDF constraints from table 5 on
the frequentist fits of data sets A and B. In particular for each
value of $r=g_{Z^\prime}/g_Z$ we display $m_H(95\%)$, defined as in
table 8, as the largest value of \mhsp on the 90\% contours (figures 3
- 7) consistent with the corresponding upper limit on \txsp from table
5. There is no CDF constraint for $Q_X= T_3R$.

\begin{table}
\begin{center}
\vskip 12pt
\begin{tabular}{c|c||c|c||c|c}
\hline
\hline
 &   & \multicolumn{2}{c}{{\bf Data Set A}}& 
                \multicolumn{2}{c}{{\bf Data Set B}}\\
Model &r & $m_H(95\%)$ CDF &$ m_H(95\%)$ LEP II & $m_H(95\%)$ CDF 
    &$ m_H(95\%)$ LEP II  \\
\hline
\hline
             & 0.27 & 230 &      & 260 &     \\
$\theta_X=0$ & 0.13 & 225 & 230  & 190 & 260 \\
             & 0.081& 206 &      & 139 &     \\
\hline
             &0.20  &220  &  & 224 & \\
$\pi/6$      &0.098  &207  &220  &163  &252 \\  
             & 0.059 &194  &  &143  & \\
\hline
             & 0.20 &214  &  &172  & \\
$\pi/3$      & 0.098 &193  &214  &141  &221 \\  
             &0.059  &188  &  &134  & \\
\hline
             &0.24  &188  &  &134  & \\
$11\pi/24$      & 0.12 &183  &202  &127  &158 \\  
             &0.072  &181  &  &124  & \\
\hline
\hline
\end{tabular}
\end{center}
\caption
{Effect of CDF bounds on the Higgs boson mass from 
frequentist fits of data sets A and B. As in 
table 8, $m_H(95\%)$ is the maximum value of \mhsp 
on the 90\% frequentist contours (figures 3 - 7) that is consistent with the 
CDF direct limit on \txsp for given values of $r=g_Z/g_{Z^\prime}$. The 
values of $m_H(95\%)$ required by the LEP II bounds on 
\tx, which are independent of $r$, are shown for comparison.}
\end{table}

\noindent {\it 5b. Bayesian Fits}

In the frequentist fits presented above we scanned over \mhsp as a
free parameter, with no prior assumption except the exceedingly mild
prior, $10 < m_H < 3000$ GeV, that was used only to obtain the
confidence levels in table 8 for the regions in the $m_H,T_X$ plane
allowed by the direct LEP II limits on \mhsp and \gzp.  The EW
precision data alone determines the outcomes of those fits, which make
predictions about the value of \mhsp that can be tested for
consistency with the direct LEP II lower bound on \mh. In this section
we follow a different procedure: we suppose that the Higgs boson has
been discovered at a specific mass which is imposed as a prior
constraint on the fits and ask how well the models describe the
precision data for that value of \mh. This is the approach followed in
\cite{flv}. We refer to this procedure as Bayesian because it assumes
a prior value for \mh.

In table 10 we present results for $m_H= 114, 225$, and 300 GeV.
Since \mhsp is fixed these fits have one more degree of freedom than
the corresponding fits in section 5a. For each fit we present the
minimum $\chi^2$, the corresponding confidence level, the change
in \chisqsp relative to the SM, and the value of \txsp at the \chisqsp
minimum. When \txsp at the \chisqsp minimum exceeds the LEP II limit
tabulated in table 4, we instead evaluate the fit with \txsp set to
the limit (marked by asterisks in table 10), so that the quoted
\chisqsp is then the smallest value consistent with the LEP II limit.

\begin{table}
\begin{center}
\vskip 12pt
\begin{tabular}{c|c||c|c|c|c||c|c|c|c}
\hline
\hline
 &   & \multicolumn{4}{c}{{\bf Data Set A}}& 
                \multicolumn{4}{c}{{\bf Data Set B}}\\
$m_H$ &Model & $\chi^2/N$&$T_X$&CL &$\Delta \chi^2$ & 
              $\chi^2/N$&$T_X$&CL &$\Delta \chi^2$  \\
\hline
\hline
 114  & SM & 17.0/12 & ... & 0.15 & ... &9.10/10 & ... & 0.52 &... \\
\hline
        &$\theta_X=0$ &17.0/11 &0.003 &0.11  &0.0 &
                     6.15/9 &0.043  &0.72 &2.96 \\
        &$\pi/6$ &17.0/11 &0.003 &0.11  &0.0 &
                     5.87/9 &0.037  &0.75 &3.24 \\
        &$\pi/3$ &17.0/11 &0.002 &0.11  &0.0 &
                     5.72/9 &0.027  &0.77 &3.39 \\
        &$11\pi/24$ &17.0/11 &0.001 & 0.11 &0.0 &
                     6.32/9 &0.0059*  &0.71 &2.79 \\
        &$-\pi/4$ &17.0/11 &0.003 & 0.11 &0.0 &
                     7.48/9 &0.045  &0.59 &1.63 \\
\hline
\hline
 225 & SM & 25.0/12 & ... & 0.015 & ... & 20.5/10 & ... & 0.025 &... \\
\hline
        &$\theta_X=0$ &21.2/11&0.047&0.031&3.8&9.0/9&0.089&0.44 &11.5 \\
        &$\pi/6$ &21.4/11 &0.038 &0.029&3.6 &9.0/9&0.073&0.43 &11.5 \\
        &$\pi/3$ &21.7/11 &0.025 &0.027&3.3 &10.0/9&0.039*&0.35 &10.5 \\
        &$11\pi/24$ &22.5/11 &0.0059* &.021&2.5&13.4/9 &0.0059*&0.15&7.1 \\
        &$-\pi/4$ &21.6/11 &0.068 &0.028&3.4 &11.6/9 &0.12&0.23&8.9 \\
\hline
\hline
 300 & SM &31.8/12 & ... & 0.0015 & ... & 28.7/10 & ... & 0.0014 &... \\
\hline
        &$\theta_X=0$ &24.6/11&0.062&0.01&7.2&11.7/9&0.11&0.23&17.0 \\
        &$\pi/6$ &24.9/11&0.054&0.01&6.9&11.8/9&0.084*&0.22 &16.9 \\
        &$\pi/3$ &25.5/11&0.025&0.008&6.3&14.7/9&0.039*&0.10 &14.0 \\
        &$11\pi/24$ &27.8/11 &0.0059*&0.003&4.0&22.0/9 &0.0059*&0.01 &6.7 \\
        &$-\pi/4$ &24.8/11&0.10&0.01&7.0&14.9/9&0.15&0.09 &13.8 \\
\hline
\hline
\end{tabular}
\end{center}
\caption
{Bayesian fits of data sets A and B assuming fixed values of \mhsp at
  114, 225, and 300 GeV. \chisqsp is the chi-square minimum and $N$ is 
  the number of degrees of freedom. 
  \txsp is the value at the \chisqsp minimum
  unless it exceeds the LEP II limit in table 4, in which case the fit
  is evaluated at the LEP II limit, denoted by an asterix. CL is the
  \chisqsp confidence level and $\Delta \chi^2$ is the \chisqsp
  difference between the \zpsp model and the SM
  fit.}
\end{table}

For data set A the Bayesian \zpsp fits at $m_H=114$ GeV do not improve
on the SM fit, and the confidence levels are lower than the SM CL.
For $m_H=225$ and 300 GeV the \zpsp fits of set A have larger CL's
than SM fit but they are still unacceptably low, $\ltap\ 0.03$ and
$\ltap\ 0.01$ respectively.  For data set B the \zpsp models have a
greater effect on the fits, and in all cases they improve on the SM.
For $m_H=225$ and 300 GeV the confidence levels of the \zpsp fits are
larger than the SM CL's by one and two orders of magnitude
respectively, and the $\Delta \chi^2$ values are highly significant.
The \zpsp fit for $\theta_X = 11\pi/24$ is severely constrained by the
strong LEP II limit on \tx.  The \zpsp models with the greatest effect
on the fits are in the range $0\ \ltap\ \theta_X\ \ltap\ \pi/3$, with
the effect for $\theta_X \simeq \pi/3$ restricted by the LEP II limit
on \txsp for the larger values of \mh. For $0\ \ltap\ \theta_X \
\ltap\ \pi/6$ the confidence levels are quite acceptable all the way
up to $m_H= 300$ GeV.  The large values of $\Delta \chi^2$ in table 10
are unambiguous evidence of the effectiveness of the \zpsp model for
set B with $m_H= 225$ and 300 GeV.

Contour plots for these Bayesian \chisqsp fits are shown in figures 8
- 12. For the \zpsp models we exhibit the 90\% and 95\% contours with
the LEP II limits on \mhsp and \txsp superimposed. The 90 and 95\%
confidence intervals for the corresponding SM fits are indicated by
the tick marks on the horizontal dot-dashed line, elevated above the
x-axis for visibility. With the Bayesian prescription, these intervals
mark the value of \mhsp at which CL$(\chi^2/N)= 0.10$ or 0.05, with
$N=12$ for the SM fit to set A and $N=10$ for B.  Similarly the \zpsp
contour plots are the 90 and 95\% trajectories in the (\mh, \tx) plane
with $N= 11$ and 9 for A and B respectively.  Table 11 presents 95\%
upper limits on \mhsp from these fits, defined for the SM as the upper
limit of the 90\% symmetric Bayesian confidence interval and for the
\zpsp models as the largest value of \mhsp on the 90\% contour that is
consistent with the LEP II limit on \tx. For data set A the 95\% upper
limits of the \zpsp models are lower than for the SM, while for set B
the limits increases relative to the SM, by a factor $\simeq 2$ to
nearly 400 GeV for the $Y$-sequential boson.  A qualitatively similar
conclusion was reached by Ferroglia {\it et al.},\cite{flv} who used 
the statistical method that we refer to here as Bayesian.

It is interesting to reflect on the differences in the \mhsp
confidence intervals for the frequentist and Bayesian fits of data
sets A and B. Consider first the SM fits. In the frequentist fits the
95\% upper limit for \mhsp (the maximum of the 90\% symmetric
confidence interval) is 153 GeV for set A and 105 GeV for set B, while
for the Bayesian fits the pattern is reversed with 143 GeV for A and
183 GeV for B. The difference is due to the smaller confidence level
of the fits to set A, e.g., CL$(16.54,12) = 0.13$ for the frequentist
fit to set A, compared to CL$(5.63,9) = 0.78$ for B. The greater reach
in \mhsp of the Bayesian SM fit to set B is a consequence of the
higher confidence level of the fit at the \chisqsp minimum, which
allows for a greater excursion in \mh, even though \mhsp at the
\chisqsp minimum is smaller for B than for A. In the frequentist fits
the $\Delta \chi^2$ method is used to compute the confidence
intervals. In that method one computes the change in \chisqsp from the
\chisqsp minimum, without regard to what the value of \chisqsp
actually is at the minimum, so there is no penalty for the larger
\chisqsp minimum of set A, and the 90\% interval reaches to larger
\mhsp because of the influence of the hadronic asymmetries. In a sense 
the $\Delta \chi^2$ method is Bayesian, since it assumes the fit 
at the \chisqsp minimum as a prior and then estimates the likelihood 
for deviations from the minimum. 

\begin{figure} 
\centerline{\includegraphics[width=2.5in,angle=90]{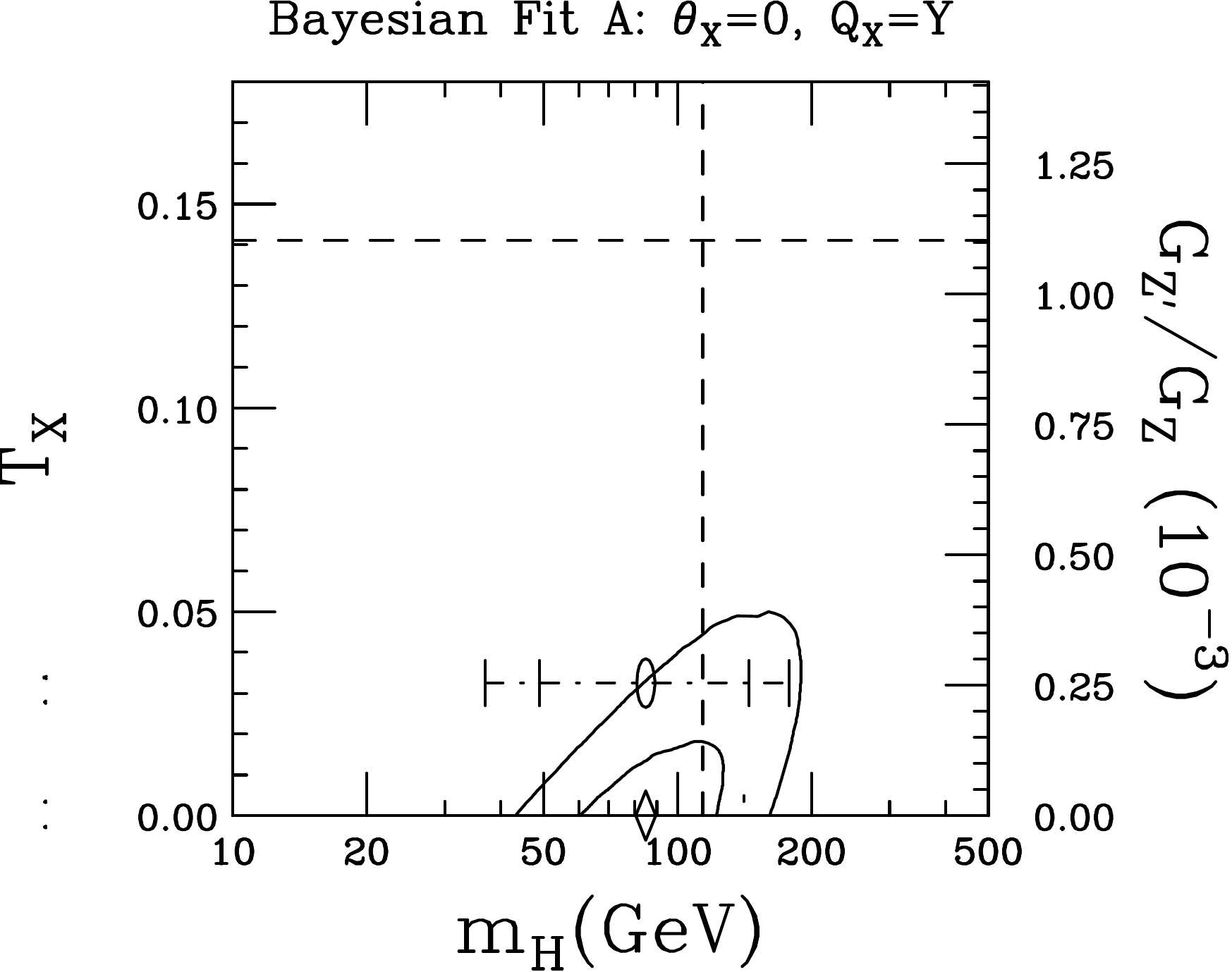}
\hspace*{.2cm}
\includegraphics[width=2.5in,angle=90]{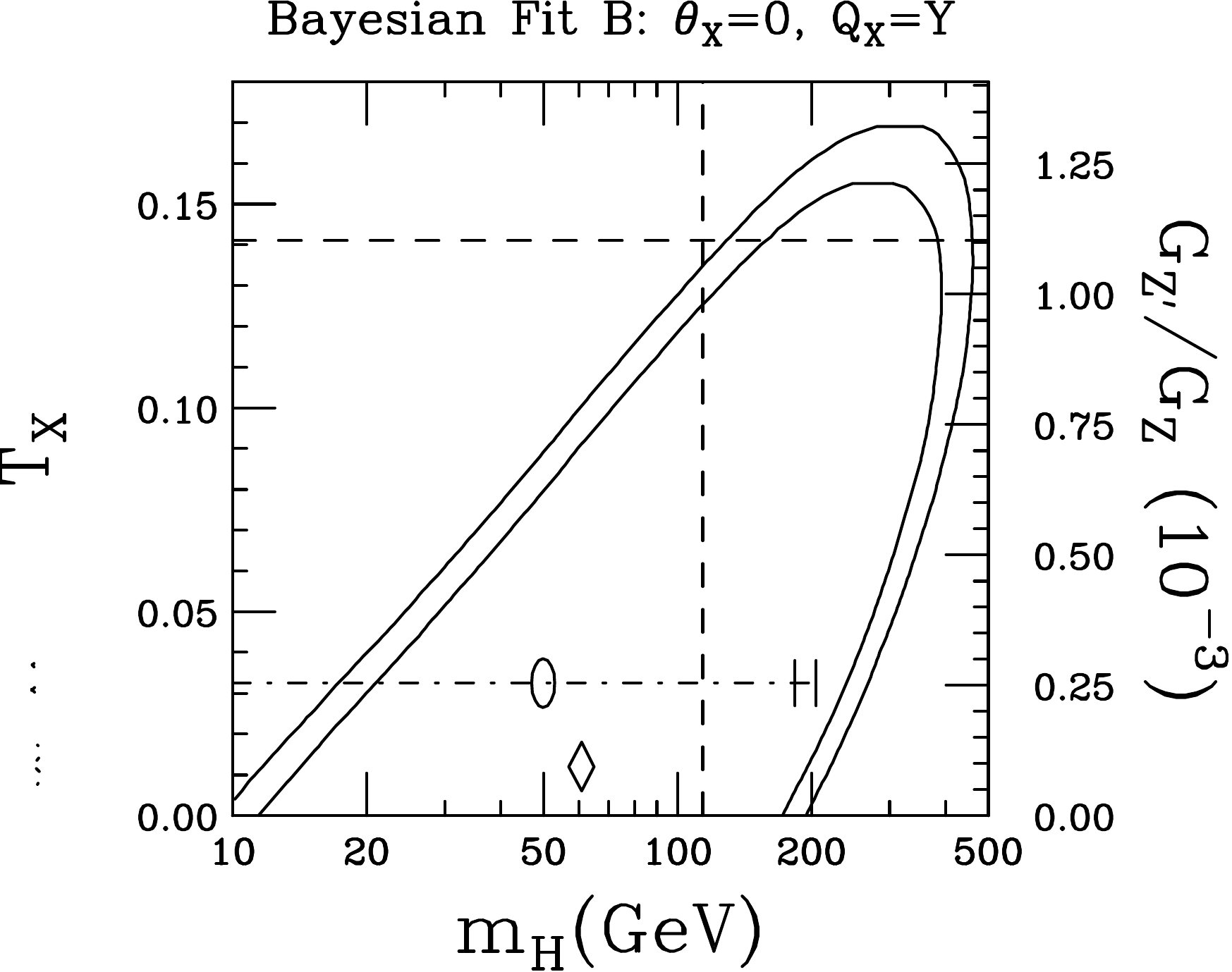}}
\caption{90\% and 95\% CL contours in the $T_X - m_H$ plane for
  Bayesian fits, as defined in the text, to data sets A
  and B for the $Y$-sequential model, $\theta_X=0$. The right axis
  indicates the corresponding values of $\hat G_{Z^{\prime}}=
  G_{Z^{\prime}}/G_Z$ per equation (20). The diamond indicates $m_H,
  T_X$ at the \chisqsp minimum for the \zpsp model. (Note that for set
  A for all $\theta_X$ the \chisqsp minimum is at $T_X =0$ and the
  diamond is hiding on the x-axis.)  The ellipse and dot-dash horizontal line
  display the central value and 90, 95\% symmetric (Bayesian)
  confidence intervals of \mhsp for the SM fit (elevated above $T_X =
  0$ only for clarity). The horizontal dashed line is the 95\% CL
  upper limit on \txsp extracted from LEP II data. The vertical
  dashed line is the LEP II 95\% lower limit on \mh.}
\label{fig8}
\end{figure}

\begin{table}
\begin{center}
\vskip 12pt
\begin{tabular}{c||c|c||c|c}
\hline
\hline
    & \multicolumn{2}{c}{{\bf Data Set A}}& 
                \multicolumn{2}{c}{{\bf Data Set B}}\\
Model & $m_H(95\%)$&$T_X$&$m_H(95\%)$&$T_X$ \\
\hline
\hline
SM & 143 GeV &...& 183 GeV & ... \\
$\theta_X = 0 $& 127&0.01&390&0.13\\
$\pi/6$&128&0.01&368&0.084*\\
$\pi/3$&128&0.007&300&0.039*\\
$11\pi/24$& 128&0.003&220&0.0059*\\
$-\pi/4$&124&0.01&300&0.13\\
\hline
\hline
\end{tabular}
\end{center}
\caption {95\% upper limits on \mhsp with the 
  corresponding value of \tx, from the Bayesian fits.  
  $m_H(95\%)$ is defined as the largest value of \mhsp on the 90\% 
  Bayesian contours (figures 8 - 12) consistent with the 
  LEP II upper limit on \tx.
  An asterix indicates that \mhsp is evaluated  
  for \txsp at the LEP II upper limit from table 4.}
\end{table}

\begin{figure}
\centerline{\includegraphics[width=2.5in,angle=90]{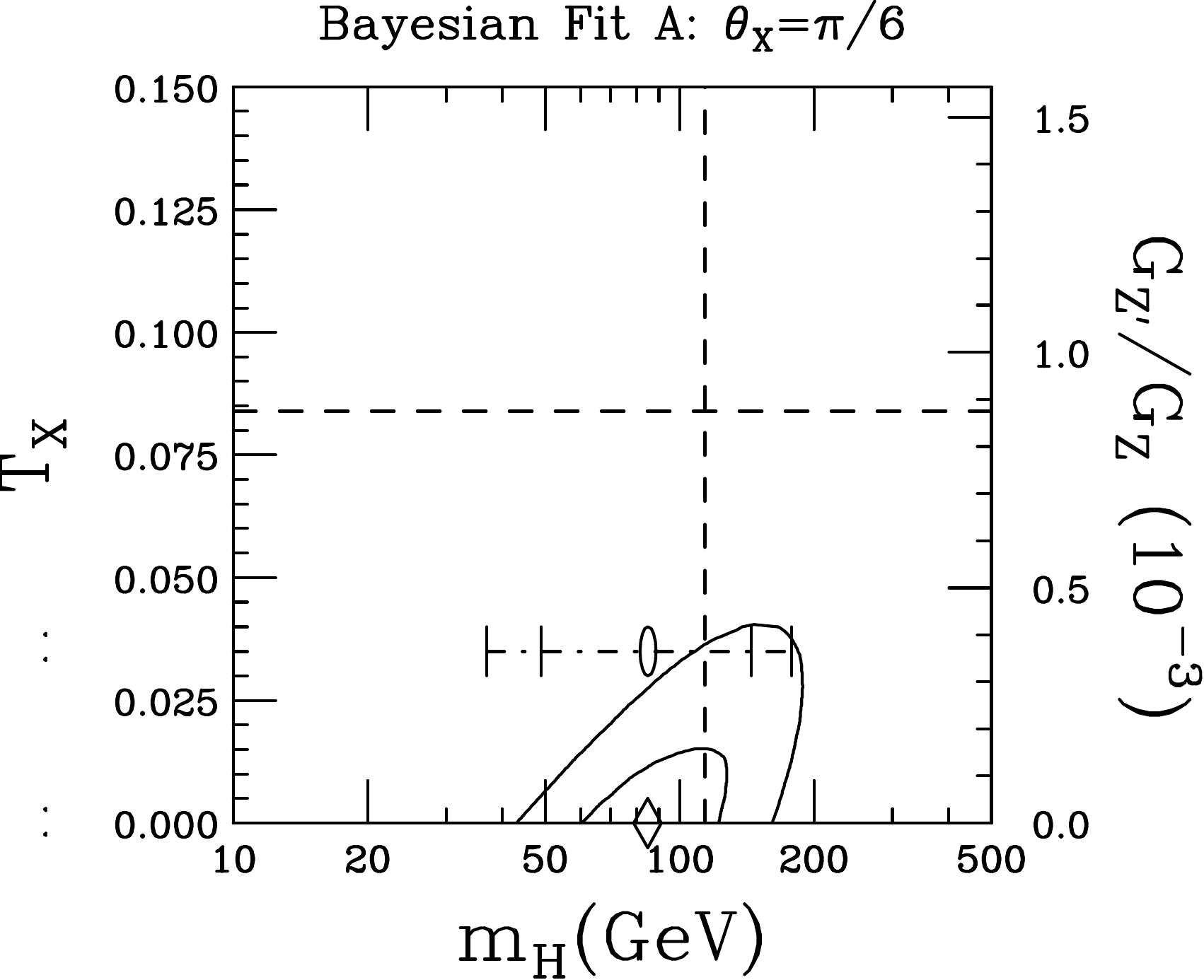}
\hspace*{.2cm}
\includegraphics[width=2.5in,angle=90]{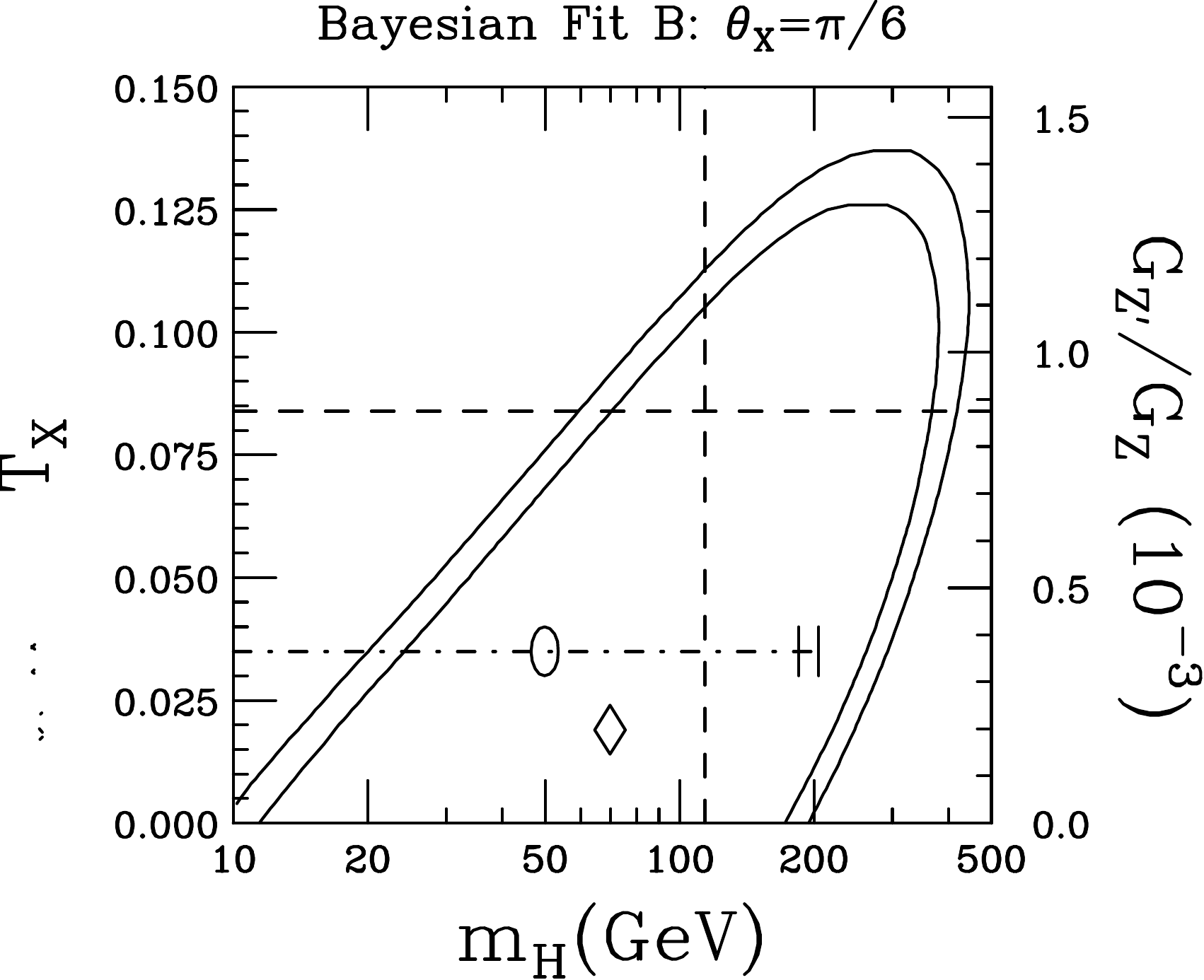}}
\caption{90\% and 95\% CL Bayesian contours for \zpsp model with 
$\theta_X=\pi/6$, as in figure 8.}
\label{fig9}
\end{figure}

\begin{figure}
\centerline{\includegraphics[width=2.5in,angle=90]{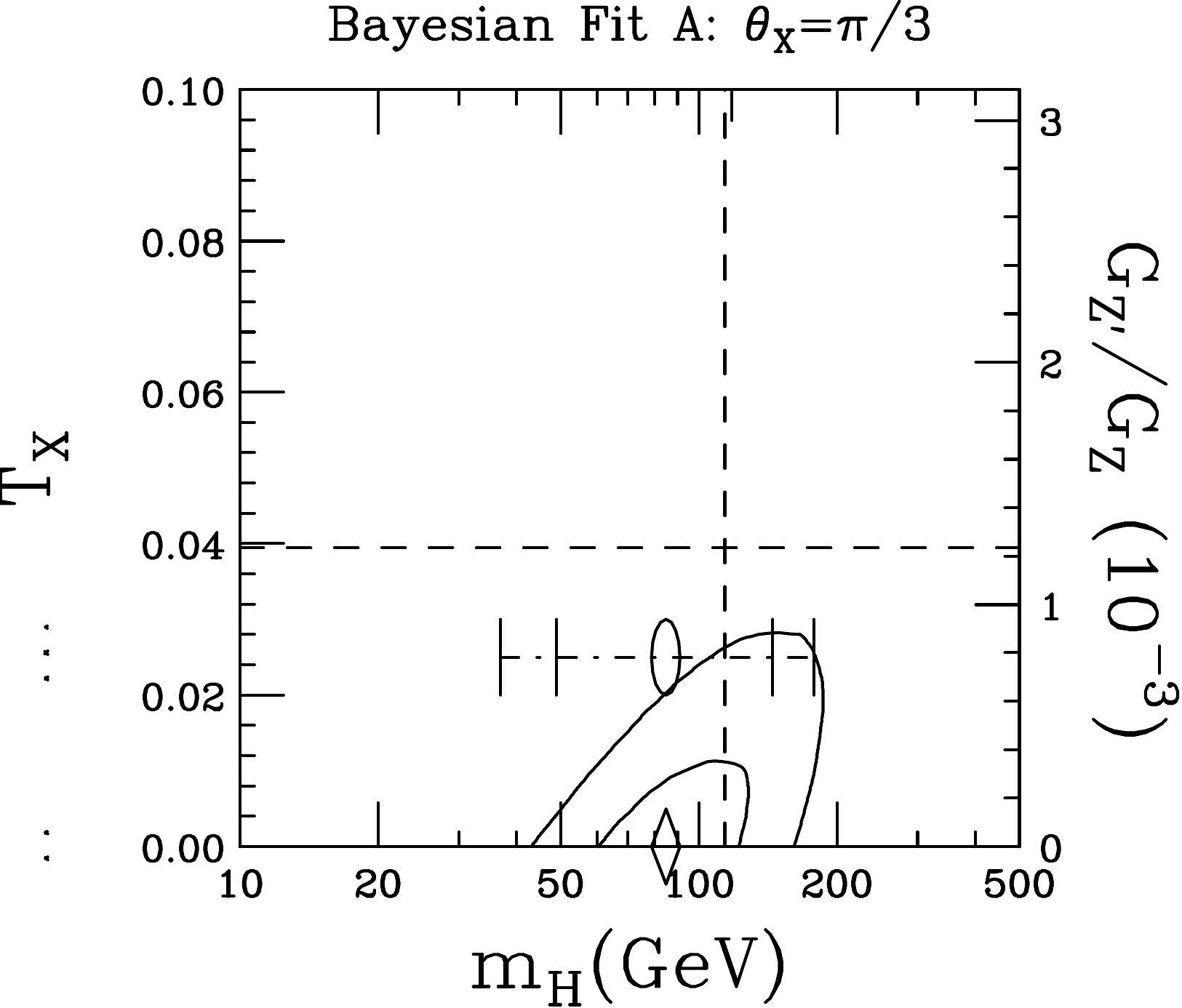}
\hspace*{.2cm}
\includegraphics[width=2.5in,angle=90]{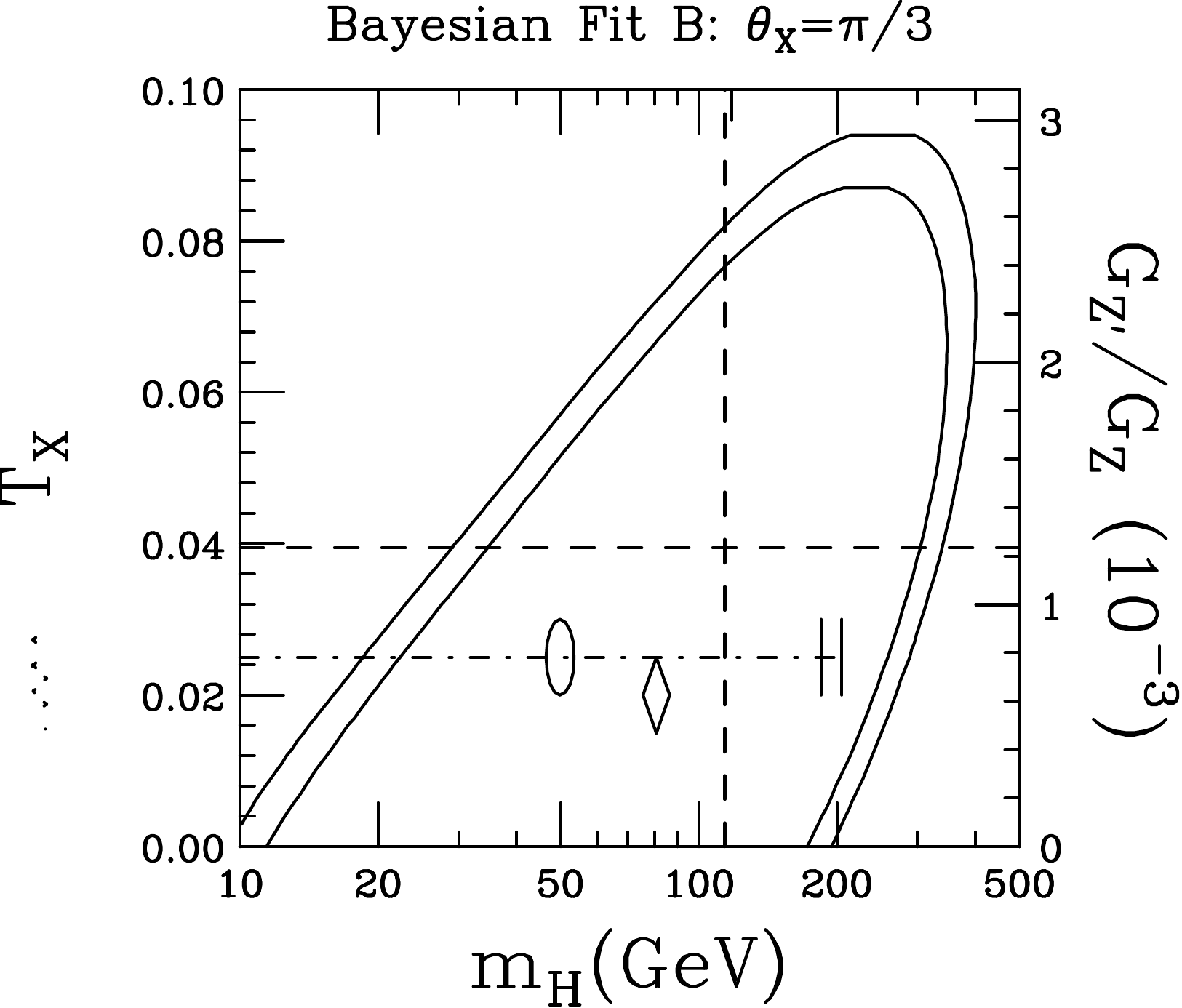}}
\caption{90\% and 95\% CL Bayesian contours for \zpsp model with 
$\theta_X=\pi/3$, as in figure 8. }
\label{fig10}
\end{figure}

\begin{figure}
\centerline{\includegraphics[width=2.5in,angle=90]{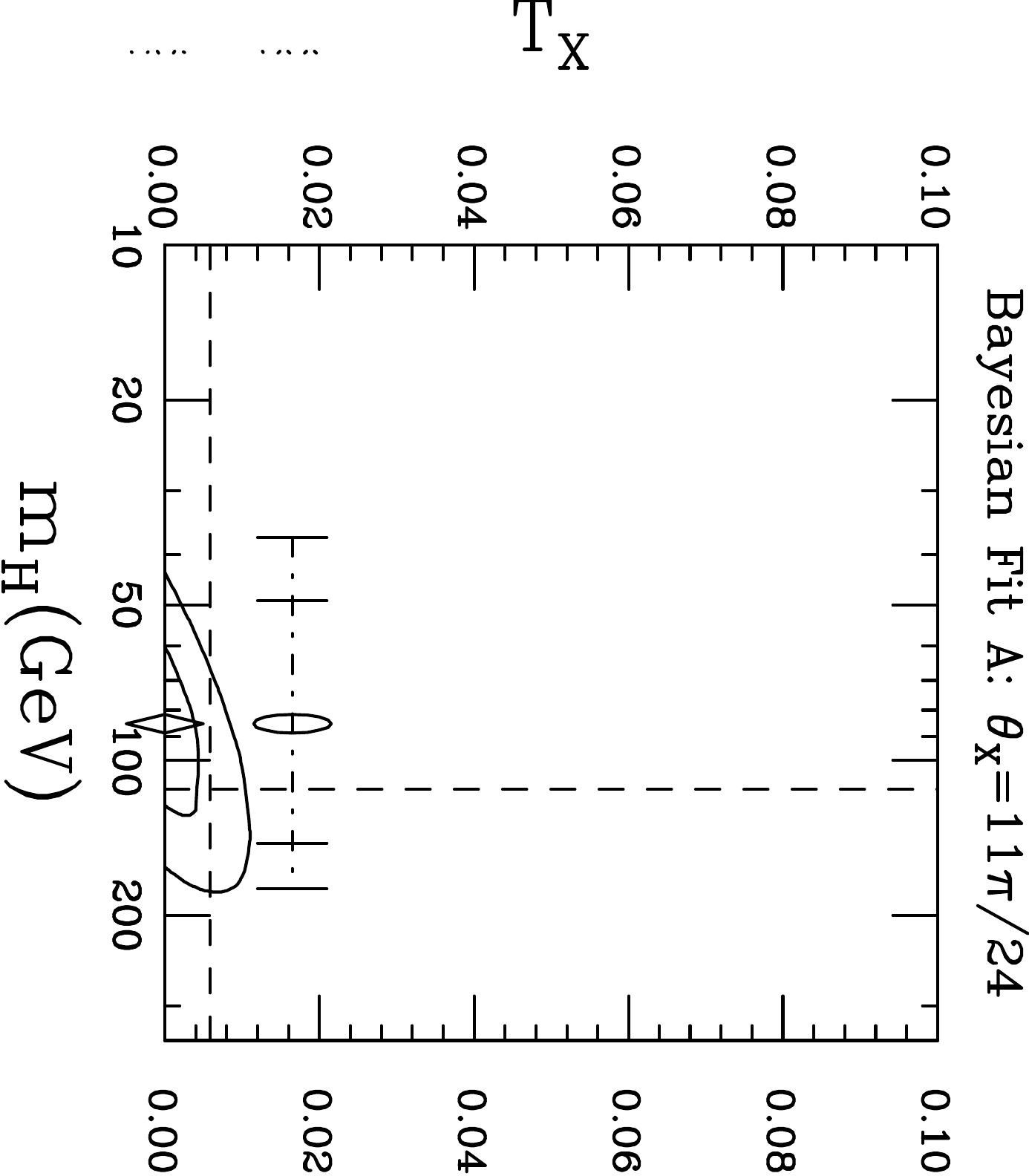}
\hspace*{.2cm}
\includegraphics[width=2.5in,angle=90]{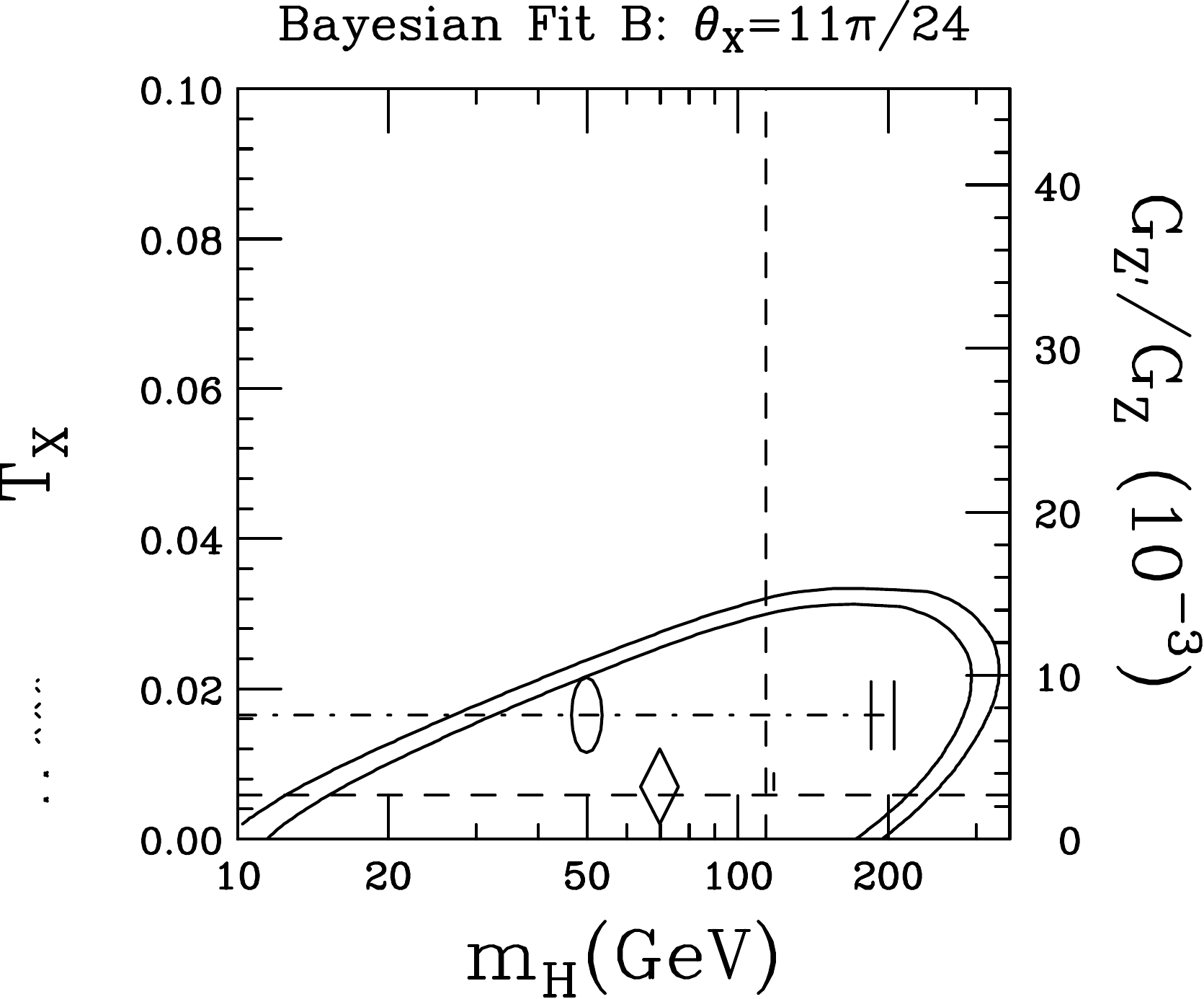}}
\caption{90\% and 95\% CL Bayesian contours for \zpsp model with 
$\theta_X=11\pi/24$, as in figure 8.}
\label{fig11}
\end{figure}

\begin{figure}
\centerline{\includegraphics[width=2.5in,angle=90]{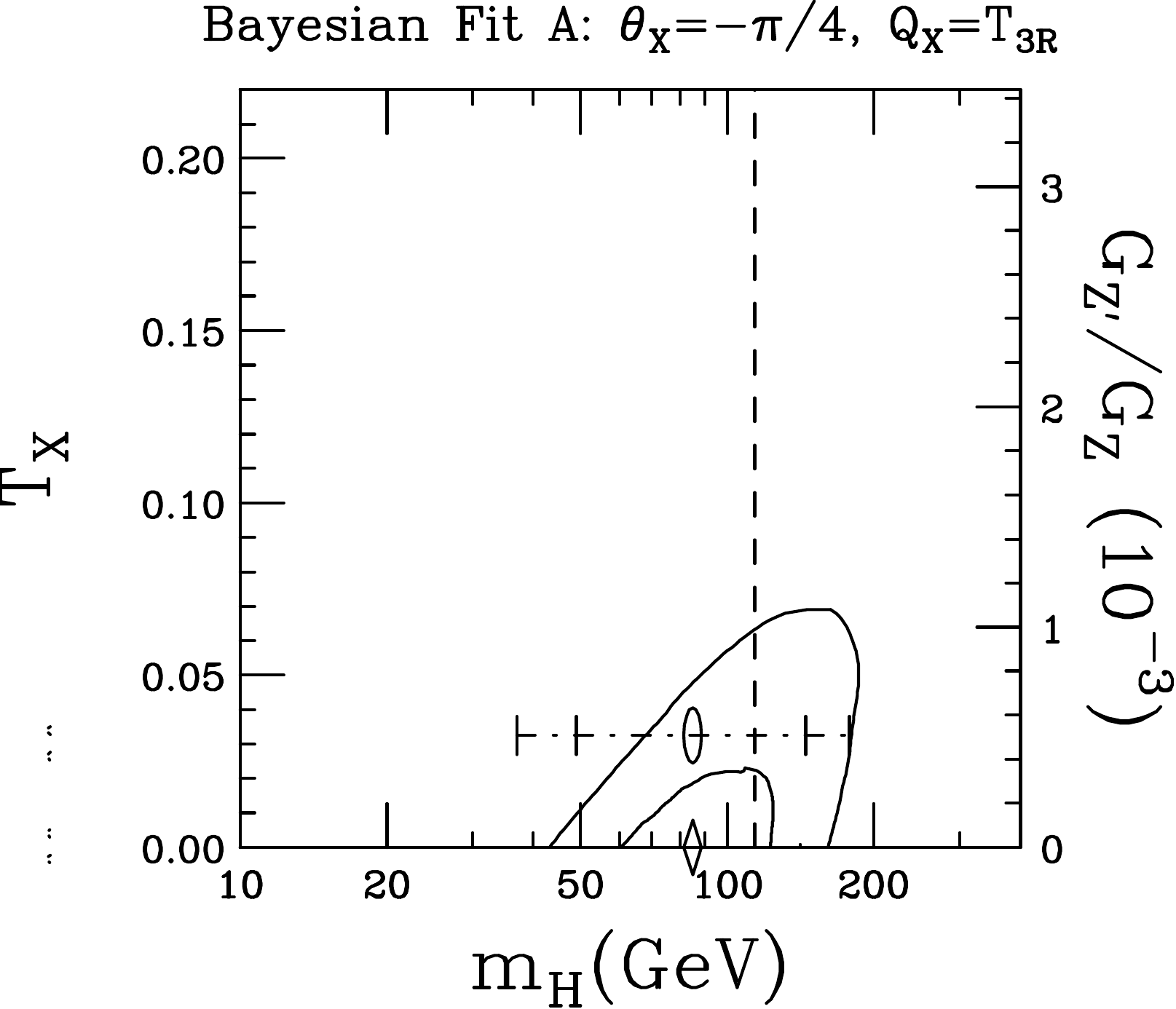}
\hspace*{.2cm}
\includegraphics[width=2.5in,angle=90]{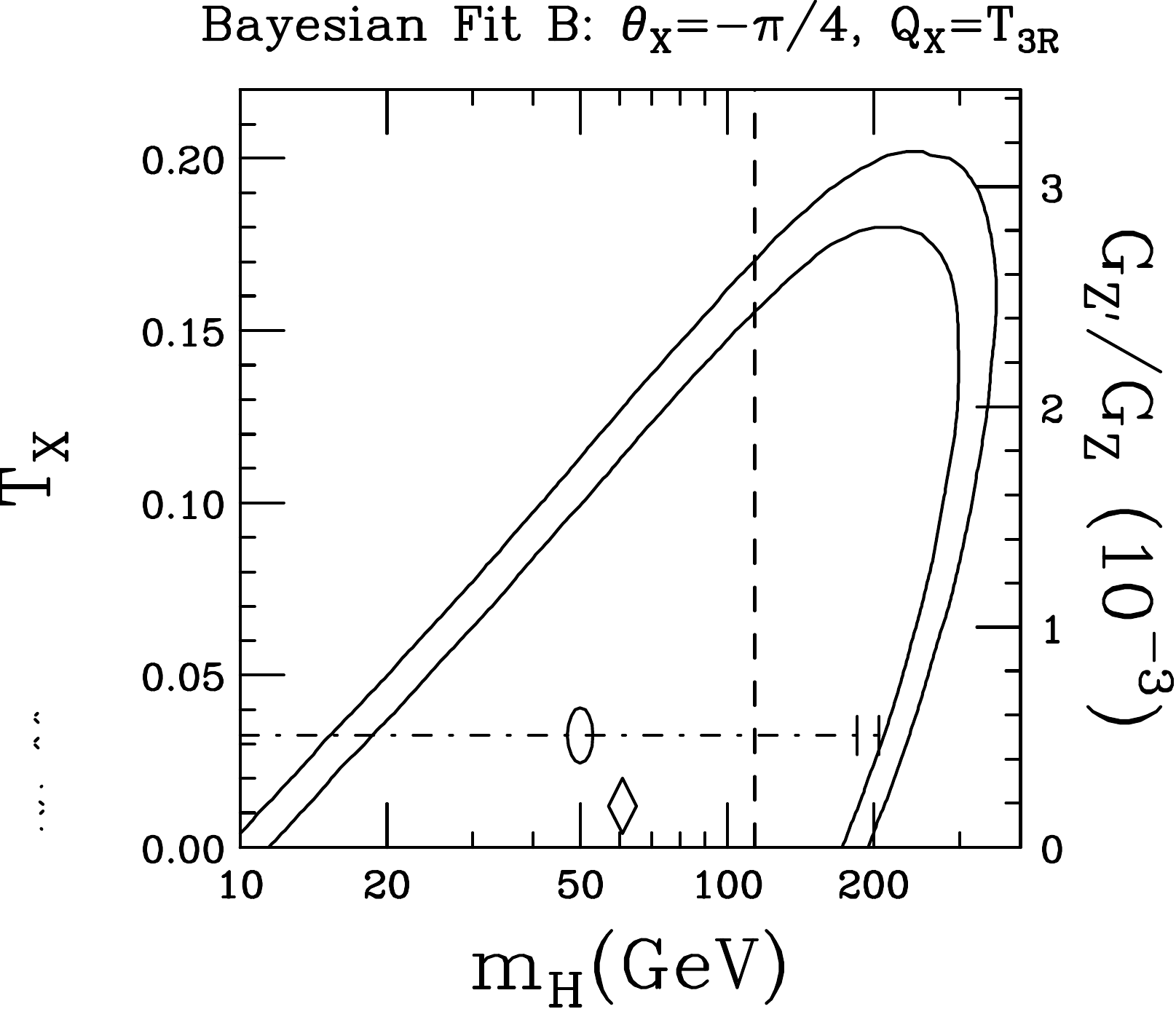}}
\caption{90\% and 95\% CL Bayesian contours for \zpsp model with 
$\theta_X=-\pi/4$, $Q_X=T_{3R}$, as in figure 8. The LEP II upper limit, 
$T_X < 0.30$, is off the graph.}
\label{fig12}
\end{figure}

For data set A with \mhsp near 114 GeV the \chisqsp minima occur at
very small values of \tx, as can be seen in table 10. The resulting
fits have essentially the same \chisqsp minima as the SM fit, and
since they have one fewer degree of freedom the CL is lower than the
SM, reaching 0.10 at a smaller value of \mh. The opposite is true of
the Bayesian fits to set B, for which the \chisqsp minima in the \zpsp
models are appreciably lower than in the SM, occuring at larger \txsp,
and the robust CL allows for larger values of \mhsp before the
\chisqsp probability falls to 0.10. In the frequentist fits of \zpsp
models using the $\Delta \chi^2$ method, \mhsp also reaches larger
values for set B than for set A, but in that case the effect is due
entirely to the improvements in the fit at larger \mhsp from \zzpsp
mixing and not at all to the more robust SM fit of set B.

The effect of the CDF bounds on the predictions of the Bayesian fits
for the Higgs boson mass is shown in table 12.
There is little effect on the fits to data set A since the bounds
on \txsp from the EW fits alone are already very strong. In the case
of data set B the CDF bounds have more impact, especially for smaller
$g_{Z^\prime}$.

\begin{table}
\begin{center}
\vskip 12pt
\begin{tabular}{c|c||c|c||c|c}
\hline
\hline
 &   & \multicolumn{2}{c}{{\bf Data Set A}}& 
                \multicolumn{2}{c}{{\bf Data Set B}}\\
Model &r & $m_H(95\%)$ CDF &$ m_H(95\%)$ LEP II & $m_H(95\%)$ CDF 
    &$ m_H(95\%)$ LEP II  \\
\hline
\hline
             & 0.27 & 127 &      & 384 &     \\
$\theta_X=0$ & 0.13 & 127 & 127  & 254 & 390 \\
             & 0.081& 127 &      & 211 &     \\
\hline
             & 0.20 &128  &  &302  & \\
$\pi/6$      & 0.098 &128  &128  &220  &368 \\  
             & 0.059 &123  &  &195  & \\
\hline
             & 0.20 &128  &  &235  & \\
$\pi/3$      & 0.098 &128  &128  &196  &300 \\  
             & 0.059 &128  &  &188  & \\
\hline
             &0.24  &128  &  &190  & \\
$11\pi/24$      & 0.12 &125  &128  &180  &220 \\  
             &0.072  &124  &  &176  & \\
\hline
\hline
\end{tabular}
\end{center}
\caption
{Effect of CDF bounds on the Higgs boson mass from 
Bayesian fits of data sets A and B. As in 
table 11, $m_H(95\%)$ is the maximum value of \mhsp 
on the 90\% Bayesian contours (figures 8 - 12) that is consistent with the 
CDF direct limit on \txsp for given values of $r=g_Z/g_{Z^\prime}$. The 
values of $m_H(95\%)$ required by the LEP II bounds on 
\tx, which are independent of $r$, are shown for comparison.
}
\end{table}

\noindent {\bf 6. Discussion}

We have explored the effect of a conservative class of
\zpsp models on the Higgs mass prediction from the EW fits,
considering both the possiblity that the discrepancy is a statistical
fluctuation and that it is the result of underestimated systematic
uncertainty.  In the first case we fitted essentially all the
precision EW data, data set A, while in the second we considered the
data without the three hadronic asymmetry measurements, data set B.
The fits show that the range of allowed values for \mhsp can be
significantly expanded into the allowed region above 114 GeV for data
set B while retaining an acceptable fit to the precision data, but for
data set A the possiblities are more restricted.  In particular,
because of the marginal confidence level of the SM fit to data set A,
the Bayesian fits of the \zpsp models allow even smaller domains for
\mhsp than the SM.

\begin{figure}
\centerline{\includegraphics[width=2.5in,angle=90]{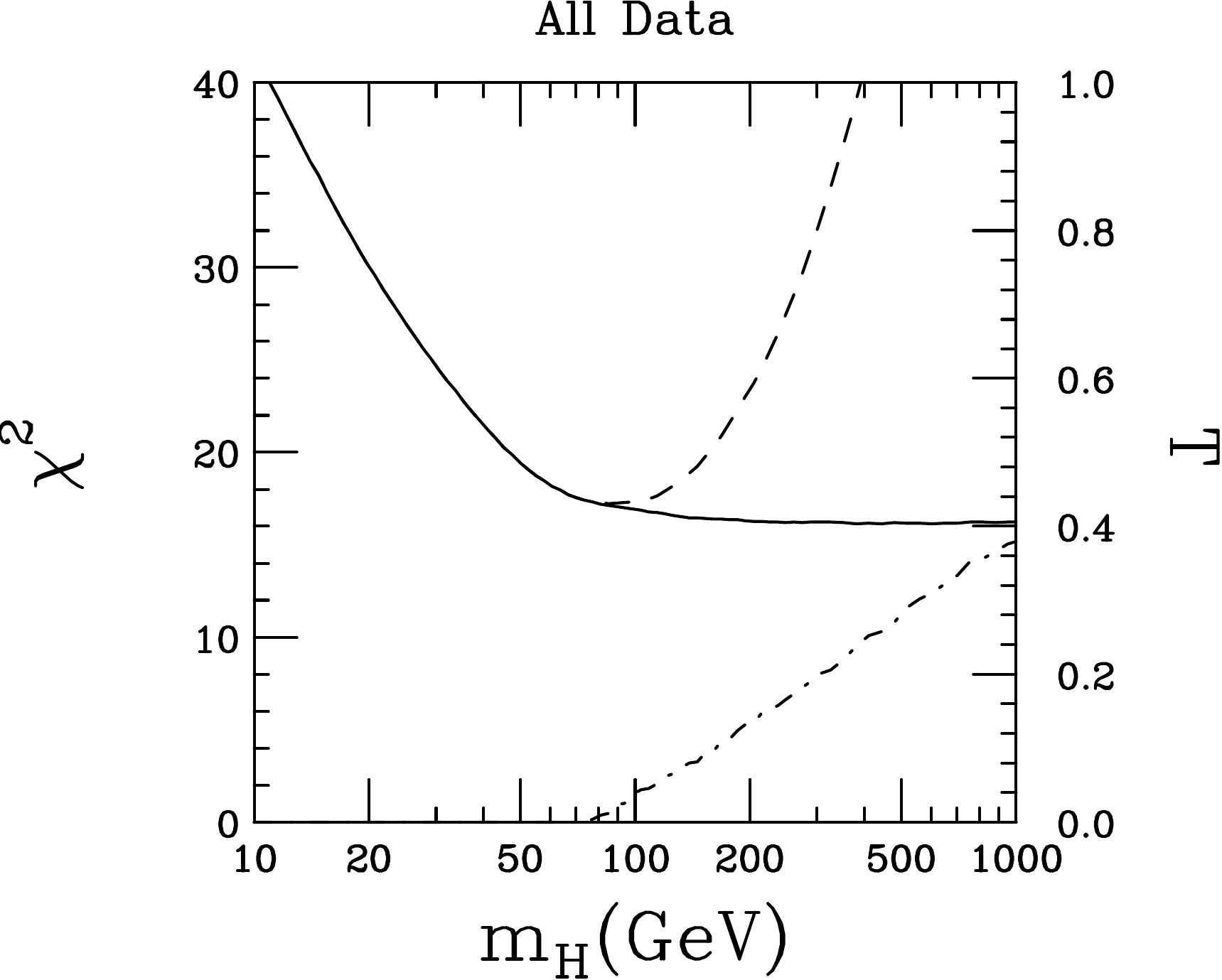}
\hspace*{.2cm}
\includegraphics[width=2.5in,angle=90]{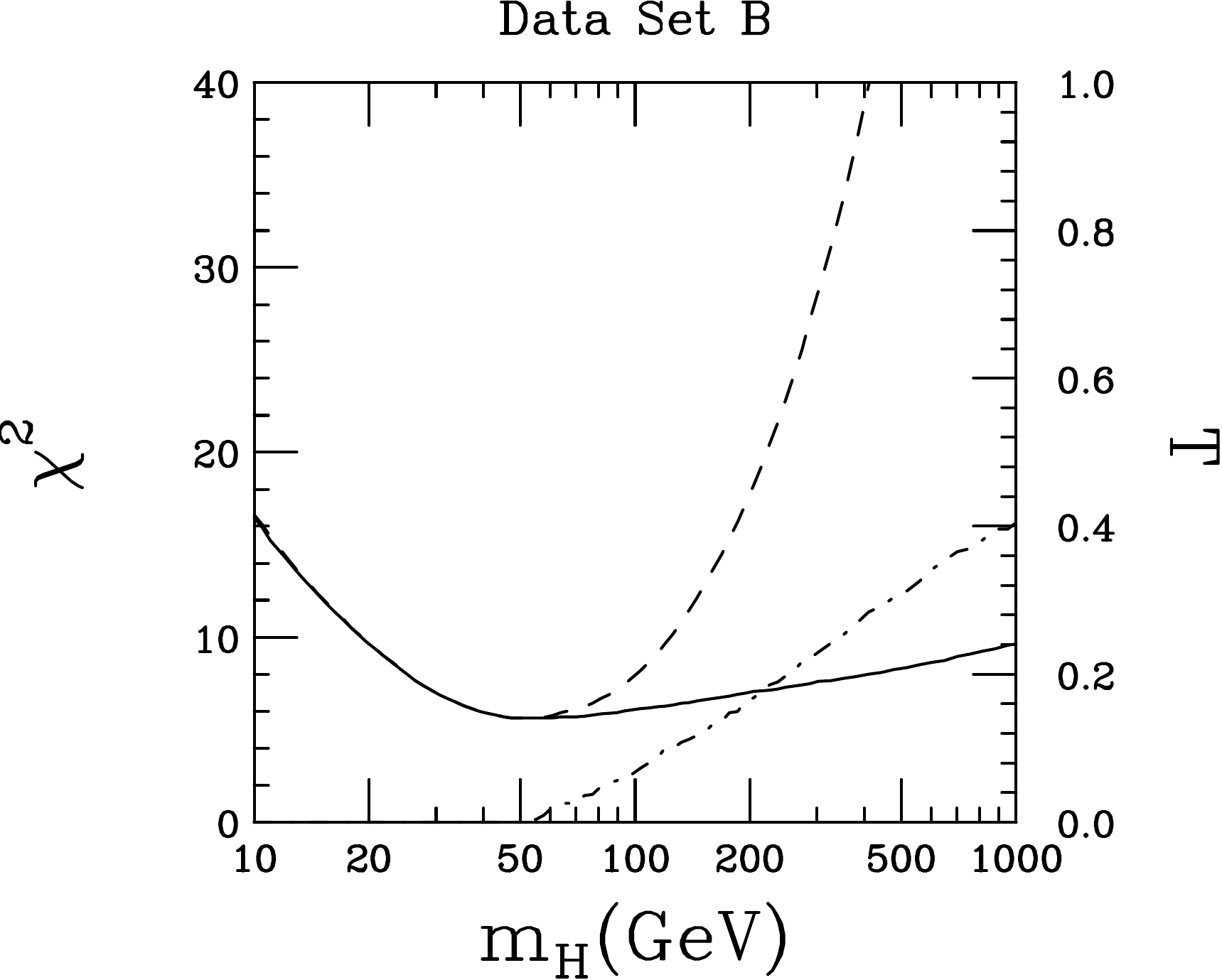}}
\caption{\chisqsp fits to the complete data set of Table 1 and to data
  set B with the oblique parameter $T>0$. The solid line is the \chisqsp
  distribution for the oblique fit, with the corresponding value of
  $T$ shown in the dot-dashed line which is read to the right axis.
  The dashed line is the \chisqsp distribution for the SM fit.}
\label{fig13}
\end{figure}

This is likely to be a generic feature of the response of the two data
sets to models of new physics, because it is typically easier to
construct models that raise the prediction for \mhsp than it is to
address the peculiarities of the \afbbsp anomaly, as would be
necessary to raise the marginal confidence level of the fit to data
set A. For instance, obliquely mediated weak isospin breaking can
raise the prediction for \mhsp toward the TeV scale without impairing
the quality of the fits but cannot improve them significantly.  Figure
13 shows that fits with the oblique parameter $T>0$, which generically
represents weak isospin breaking mediated by vacuum polarization, can
flatten the \chisqsp distribution for large values of \mh, for both
the fit to the complete data set with 12 $dof$ and for the fit to data
set B with 9 $dof$.  The \chisqsp confidence levels of these fits,
$\simeq 0.13$ for the complete set and $\simeq 0.7$ for set B, are
very near the CL's of the corresponding SM fits at their \chisqsp
minima.

Figures 14 and 15 display the 90 and 95\% confidence level contour
plots in the $m_H, T$ plane for frequentist and Bayesian 
fits. In a reversal of what we found for the \zpsp models, the
\chisqsp minimum for the all-data set is at nonzero $T$ with elevated
\mh, while for fit B it coincides with the SM fit with $T=0$ and
$m_H=50$ GeV.  However the position of the \chisqsp minima are not
very significant, since the minima are extremely shallow, as is
evident in figure 13.

\begin{figure}
\centerline{\includegraphics[width=2.5in,angle=90]{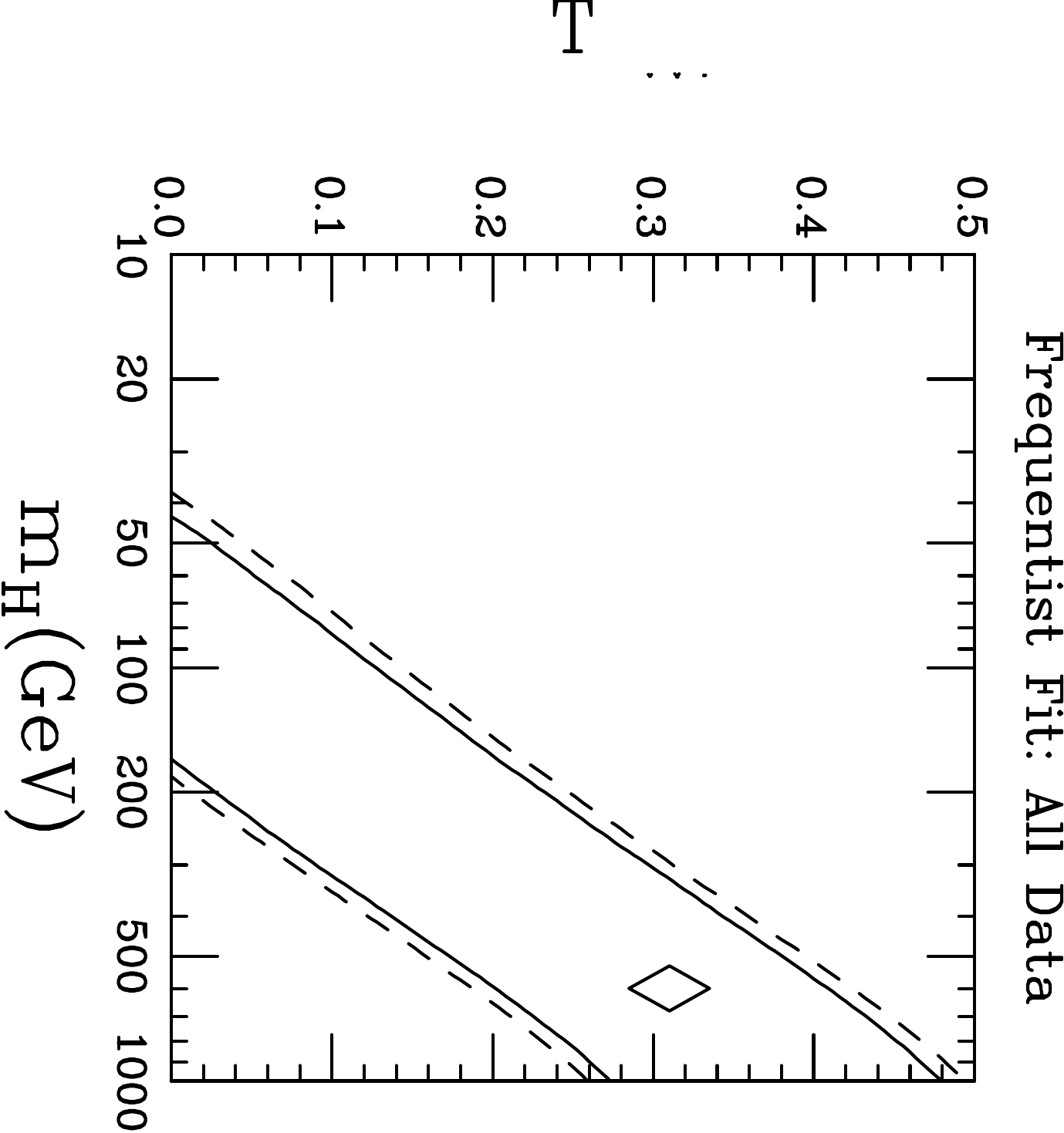}
\hspace*{.2cm}
\includegraphics[width=2.5in,angle=90]{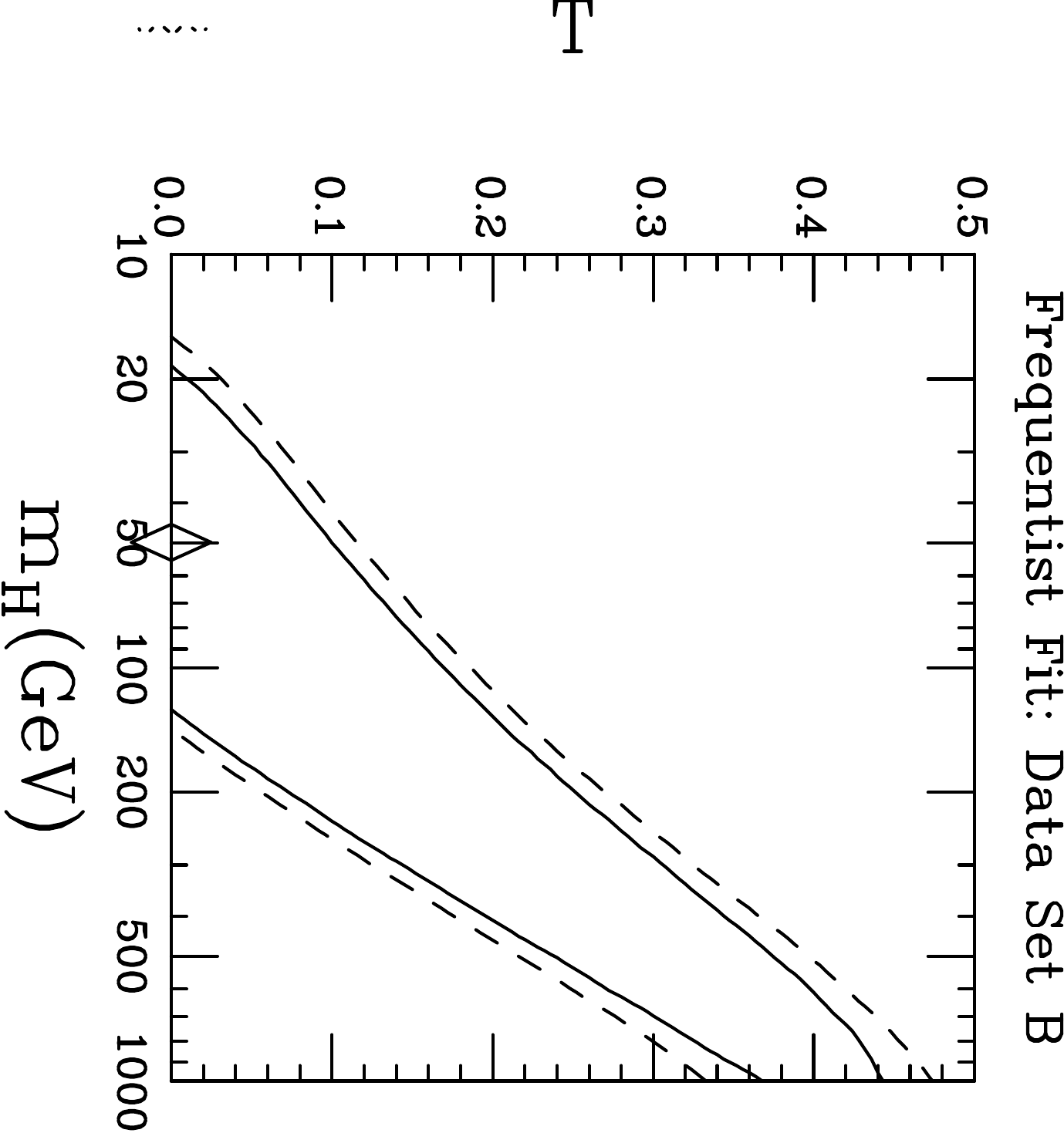}}
\caption{90 (solid line) and 95\% (dashed line) contour plots of
  frequentist fits with oblique parameter $T>0$. The
  position of the \chisqsp minimum is indicated by the diamond.}
\label{fig14}
\end{figure}

\begin{figure}
\centerline{\includegraphics[width=2.5in,angle=90]{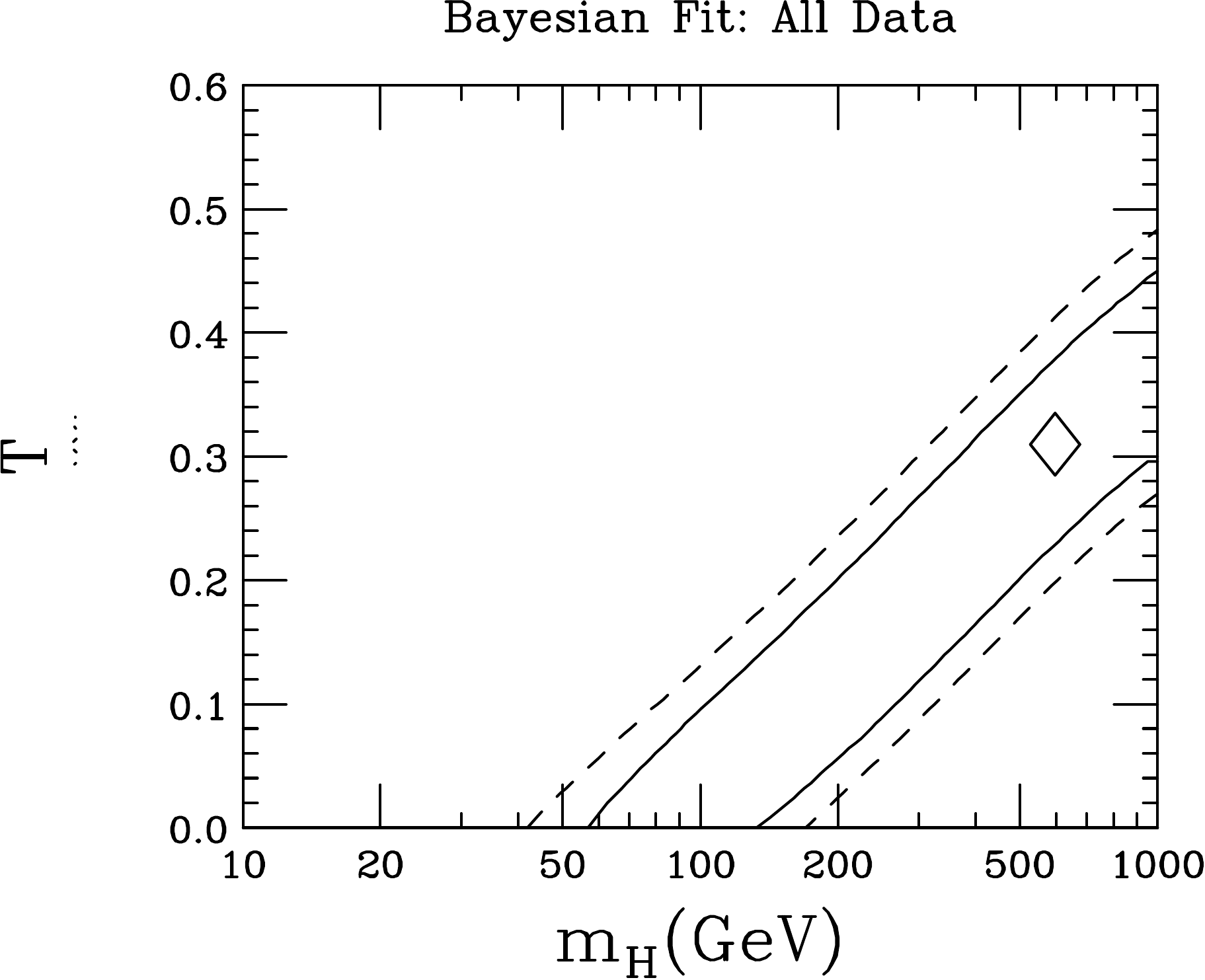}
\hspace*{.2cm}
\includegraphics[width=2.5in,angle=90]{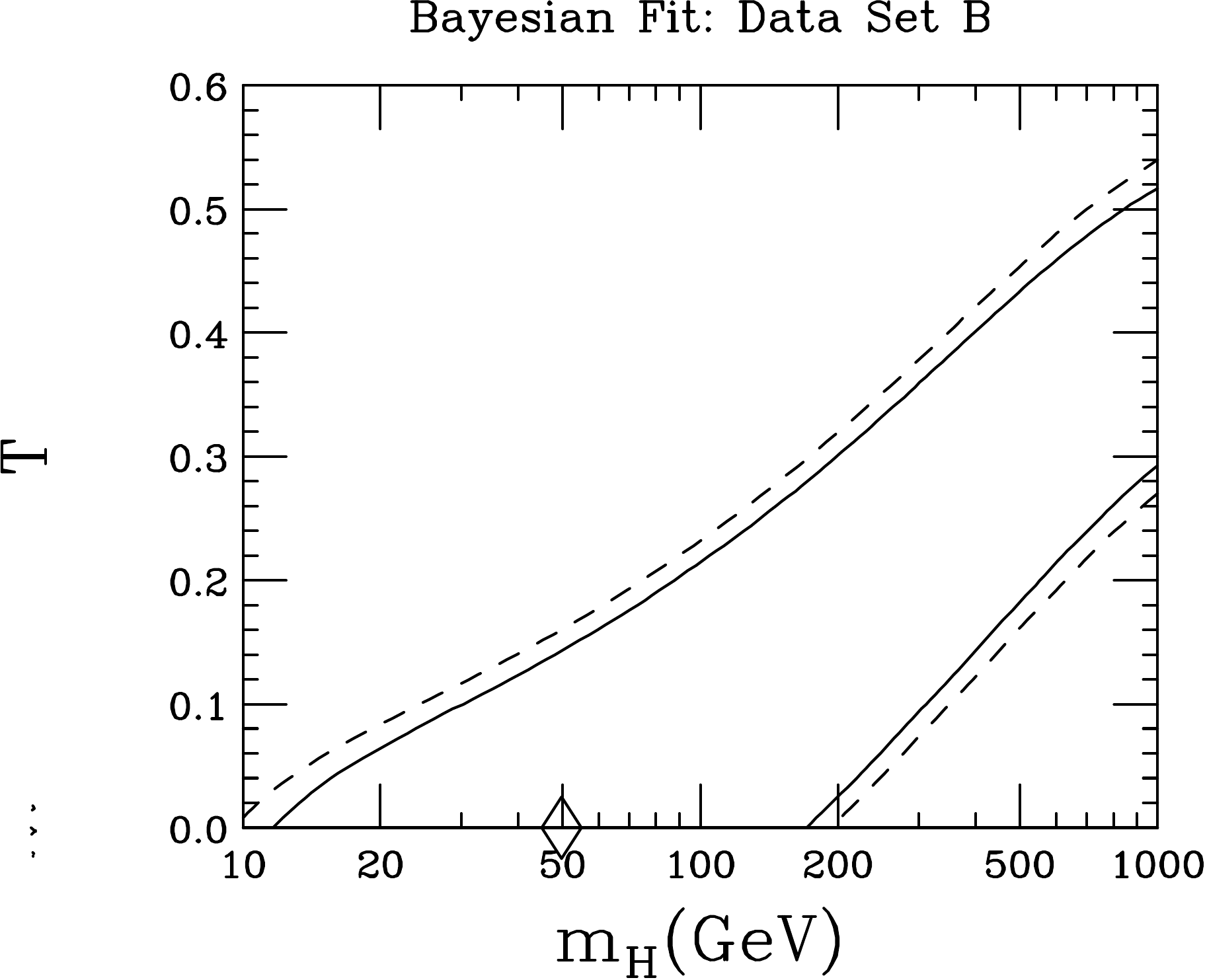}}
\caption{90 (solid line) and 95\% (dashed line) contour plots of
  Bayesian fits with oblique parameter $T>0$. The position of the
  \chisqsp minimum is indicated by the diamond.}
\label{fig15}
\end{figure}

Weak isospin breaking is the basis for the effect of \zzpsp mixing on
the \mhsp predictions presented here. In the \zpsp models the effect
is limited to $m_H\ \ltap\ 300$ GeV because the shifts in the
$Z\overline ff$ couplings are proportional to
$g_{Z^{\prime}}\theta_M$, which in turn is proportional to \tx,
$$
  g_{Z^{\prime}}\ \theta_M= g_Z{\alpha T_X\over \cos \theta_X}.
                                   \eqno{(28)}
$$
The limit on \txsp is reached when $g_{Z^{\prime}}\ \theta_M$ 
grows so large that the $Z\overline ff$ couplings deviate too 
far from their SM values, causing \chisqsp to increase.

The $3.2\sigma$ discrepancy in the SM determination of the weak mixing
angle by leptonic and hadronic asymmetry measurements sends an
ambiguous message. It reduces the confidence level of the SM fit and
raises questions about the SM prediction for \mh, which averages a
``bimodal'' distribution of measurements favoring 50 (\alr, $m_W$) and
500 GeV (\afbb), as shown in figure 1. The significance and meaning of
the discrepancy can only be clarified by future experimental data. If
for instance evidence is found for a large deviation of the
right-handed $Z\overline bb$ coupling from its SM value, it would
confirm the \afbbsp anomaly as a genuine signal of new physics. If the
discrepancy results from unresolved theoretical and/or experimental
systematic uncertainties in the very challenging hadronic asymmetry
measurements, the low SM prediction for \mhsp that results from the
remaining measurements (data set B) strongly suggests new physics to
raise the predicted value into the experimentally allowed region.

The simple class of models studied here offers a paradigm for how we
can use the LHC together with the precision EW data to understand the
underlying physics. If a \zpsp boson is discovered at the LHC, it will
be important to compare its properties as measured at the LHC with the
constraints of the EW fits. It would first be essential to study the
leptonic and hadronic couplings of the \zpsp to determine whether it
is in fact in the class of \zpsp bosons considered here and to measure
the parameter $\theta_X$. If so, the measurement of the coupling
constant $g_{Z^{\prime}}$ and mass $m_{Z^{\prime}}$ would determine
the effective Fermi constant \gzp, which in turn specifies the oblique
parameter \txsp that determines the \zpsp fits.  With the \zpsp
parameters known, the EW fit would make a prediction for the Higgs
boson mass which could be compared with direct measurement of \mhsp at
the LHC. It is then important to ascertain how well such a program
could be carried out at the LHC,\cite{zp-lhc} both in its original
incarnation and after possible luminosity and energy upgrades.

This example illustrates the important role that the precision EW data
can continue to play in the future. At the end of the day, when new
physics has been discovered and studied at the LHC, we will want to
consider how it affects the EW fit. A consistent explanation of both
the high energy data and the precision EW data would be a powerful
confirmation of the theoretical picture, just as high energy data
together with the precision data have confirmed the SM as the correct
zero'th order model. If the model used to describe the high energy
measurements is not consistent with the precision EW data, it could
mean that the model is wrong or that there is other undiscovered new
physics affecting the EW observables. 

To realize the potential of such a program, combining both the high
energy measurements and the low energy precision data, it is important
to resolve the ambiguity that the \afbbsp anomaly casts over the
current data.  Future high statistics studies at a high intensity Z
factory like the proposed Giga-Z project\cite{giga-z} could determine
if the anomaly is a statistical fluctuation and would allow further
study of the experimental systematic uncertainties. Additional work on
systematic uncertainties with a theoretical component would also be
essential, for instance, the merging of the radiative corrections
to $Z \to \overline bb$ with the experimental acceptance, which 
gives rise to a systematic uncertainty that is now very difficult to 
quantify. 

It is also possible that the LHC could illuminate the issue. For
instance, in the framework of the models discussed in this paper, the
discovery of a $Y$-sequential \zpsp boson together with a 300 GeV
Higgs boson would be compatible with the Bayesian fit to data set B
but not to data set A. In general, discoveries at the LHC could favor
a model that is consistent with one data set but not the other. The EW
fit of the compatible data set could then be compared with the direct
observations at the LHC to further constrain and test the model. In
principle, if the ambiguities can be resolved and the precision can be
improved, the EW fit could even be used to probe for additional new
physics before it is directly observed, just as the radiative
corrections to the rho parameter\cite{rho} enabled the prediction of
the top quark mass scale before the top quark was discovered.

\bigskip

\noindent {\bf Appendix: ``Pseudo-Oblique'' Parameterization of the
  $Y$-Sequential Model}

We will show, as discussed in section 3, that the effect of mixing
with $Y$-Sequential \zpsp bosons on the EW fits discussed here can be
represented by oblique parameters $S^{\prime},T^{\prime}$, given by 
$$
T^{\prime}= -T_X           \eqno{(A1)}
$$
$$
S^{\prime}= -4(1-x_W)T_X           \eqno{(A2)}
$$
where from equation (17) with $\theta_X=0$
$$
\alpha T_X = \epsilon= -{\delta m_Z^2 \over m_Z^2}. \eqno{(A3)}
$$
The precision EW fit for the $Y$-sequential \zpsp
boson model can then be extracted from the usual oblique fit by
considering the line $S= 4(1-x_W)T$ in the $S,T$ plane. 

Using equations (12 - 16) with $\theta_X=0$, the $Z\overline ff$
interaction, equation (11), is
$$
{\cal L}_f= g_Z \left(1 + {\alpha T_X \over 2}\right) 
            \overline f {\not\! Z} 
            (t_{3L}^f -q^f \hat{x}_W +\epsilon {y \over 2})f.
	                         \eqno{(A4)}
$$
We will show that the interaction can also be represented by the
obliquely corrected SM Lagrangian
$$
{\cal L}_f= g_Z \left(1 + {\alpha T^{\prime} \over 2}\right) 
            \overline f {\not\! Z} 
            (t_{3L}^f -q^f x^{\prime}_W)f.
	                         \eqno{(A5)}
$$
where $x^{\prime}_W$ has the usual oblique correction in terms of \spsp 
and \tp,
$$
x^{\prime}_W - x_W= {\alpha\over 1-2x_W}\left( 
               {S^{\prime} \over 4} - x_W(1-x_W)T^{\prime}
               \right )           \eqno{(A6)}
$$
and the $W$ boson mass also gets the ususal correction, 
$$
{\delta m_W^2 \over m_W^2}= {\alpha\over 1-2x_W}\left( 
               -{S^{\prime} \over 2} + (1-x_W)T^{\prime}
               \right ).          \eqno{(A7)}
$$

To obtain equations (A5) and (A6) we substitute $y=2(q-t_3)$ 
in equation (A4). The result to order O($\epsilon$) is
$$
{\cal L}_f= g_Z \left(1 + {\alpha T_X \over 2} -\epsilon\right) 
            \overline f {\not\! Z} 
            [t_{3L}^f -q^f (\hat{x}_W -\epsilon(1- x_W))]f.
	                         \eqno{(A8)}
$$
Matching the prefactors of equations (A5) and (A8)  
implies that $\alpha T^{\prime}= \alpha T_X - 2\epsilon = -\alpha
T_X$, which establishes equation (A1). \spsp is then fixed by the
remaining condition for the equivalence of equations (A5) and (A8), 
$x^{\prime}_W = \hat{x}_W -\epsilon(1-x_W)$, which implies
$$
{\alpha\over 1-2x_W}\left( 
               {S^{\prime} \over 4} - x_W(1-x_W)T^{\prime}
               \right ) = -\alpha T_X \left({x_W(1-x_W)\over 1-2x_W} 
                     + 1-x_W \right)     \eqno{(A9)}
$$
and using equation (A1) yields the result, equation (A2), 
for \sp.      

The corrections to the $W$ boson mass now provide a nontrivial test of
the equivalence of the \sp, \tpsp oblique representation 
with the original \zpsp model. From the original model the
correction is 
$$
{\delta m_W^2 \over m_W^2}= {1-x_w\over 1-2x_w}\, \alpha T_X. 
      \eqno{(A10)}
$$ 
Substituting the expressions for \spsp and \tpsp in terms of $T_X$
from equations (A1) and (A2) into the generic expression for $\delta
m_W$, equation (A7), the result is precisely equation (A10).  For all
other observables we consider (see table 1) the oblique corrections
enter via the weak mixing angle $x_W$ or, in the case of the $Z$ width
$\Gamma_Z$, via the prefactor $1 + \alpha T$. We are then guaranteed
that oblique fits with the constraint $S^{\prime}=4(1-x_w)T^{\prime}$
are equivalent to the fits of the original $Y$-sequential \zpsp model.
We have also verified the equivalence numerically by fitting
the data using both representations.

\vskip 0.2in

\noindent{\small This work was supported in part by the Director, Office of
Science, Office of High Energy and Nuclear Physics, Division of High
Energy Physics, of the U.S. Department of Energy under Contract
DE-AC02-05CH11231}



\end{document}